\documentclass[fleqn,usenatbib]{mnras}

\usepackage{newtxtext,newtxmath}
\usepackage[T1]{fontenc}

\DeclareRobustCommand{\VAN}[3]{#2}
\let\VANthebibliography\thebibliography
\def\thebibliography{\DeclareRobustCommand{\VAN}[3]{##3}\VANthebibliography}

\usepackage{graphicx}	% Including figure files
\usepackage{amsmath}	% Advanced maths commands
\usepackage{pdflscape}
\usepackage{subfigure}
\usepackage{chemformula}
\title[Clouds and the radius evolution of giant planets]{The effect of cloudy atmospheres on the thermal evolution of warm giant planets from an interior modelling perspective}

\author[A.J. Poser et al.]{
A. J. Poser$^{1}$,\thanks{E-mail: anna.poser@uni-rostock.de}
R. Redmer$^{1}$
\\
$^{1}$University of Rostock, Institute of Physics, D-18059 Rostock, Germany\\
}

\date{Accepted 2024 February 28. Received 2024 February 21; in original form 2023 September 18}

\pubyear{2024}

\begin{document}
\pagerange{\pageref{firstpage}--\pageref{lastpage}}
\maketitle

\begin{abstract}
We are interested in the influence of cloudy atmospheres on the thermal radius evolution of warm exoplanets from an interior modelling perspective. By applying a physically motivated but simple parameterized cloud model, we obtain the atmospheric $P$-$T$ structure that is connected to the adiabatic interior at the self-consistently calculated radiative-convective boundary.  We investigate the impact of cloud gradients, with the possibility of inhibiting superadiabatic clouds. Furthermore, we explore the impact on the radius evolution for a cloud base fixed at a certain pressure versus a subsiding cloud base during the planets' thermal evolution. 
We find that deep clouds clearly alter the evolution tracks of warm giants, leading to either slower/faster cooling than in the cloudless case (depending on the cloud model used).
When comparing the fixed versus dynamic cloud base during evolution, we see an enhanced behaviour resulting in a faster or slower cooling in the case of the dynamic cloud base. 
We show that atmospheric models including deep clouds can lead to degeneracy in predicting the bulk metallicity of planets, $Z_\mathrm{P}$. For WASP-10b, we find a possible span of $\approx {Z_\mathrm{P}}_{-0.06}^{+0.10}$. For TOI-1268b, it is $\approx {Z_\mathrm{P}}_{-0.05}^{+0.10}$. Further work on cloud properties during the long-term evolution of gas giants is needed to better estimate the influence on the radius evolution.  
 
\end{abstract}

\begin{keywords}
planets and satellites: gaseous planets -- planets and satellites: atmospheres -- planets and satellites: interiors -- planets and satellites: individual (WASP-10b, TOI-1268b) 
\end{keywords}

%%%%%%%%%%%%%%%%%%%%%%%%%%%%%%%%%%%%%%%%%%%%%%%%%%
%%%%%%%%%%%%%%%%% BODY OF PAPER %%%%%%%%%%%%%%%%%%
%%%%%%%%%%%%%%%%%%%%%%%%%%%%%%%%%%%%%%%%%%%%%%%%%%

\section{Introduction}
\label{sec:introduction}
Giant planets are essential for understanding how planets form and evolve because they hide important information within their interiors~\citep[e.g.][]{Turrini2018, Helled+21_ARIEL}. In particular, both the mass of the heavy elements and their distribution within the planet are of interest, as the bulk and atmospheric composition of the planets are related to their formation and evolutionary history. For example, the heavy element content of the planet may be correlated with the star's metallicity, both forming from the same protostellar cloud~\citep[e.g.][]{Guillot2006, Thorngren2016}.\\  
Characterisation of the interior is initially based on observational parameters of the planet, such as planetary mass, radius, and stellar age, as a proxy for the age of the planetary system and the planet itself. The ensuing description of the current (today's) bulk structure then relies on purely numerical models and delivers, among other things, the desired heavy-element content. Making use of the full set of observational parameters, one has to couple atmosphere, interior, and thermal evolution models~\citep[e.g.][]{Fortney2007, Baraffe2008, Muller2023a, Muller2023, Poser2019}.\\
In this context, the atmosphere of the planet plays a unique role. It is critical for the planet's radiative budget, in particular for irradiated planets, as it serves as a bottleneck for both the incoming stellar irradiation and the emitted intrinsic flux. As a result, it has a direct impact on the planet's cooling behaviour because the intrinsic heat from the inside is radiated away through the atmosphere over time.\\
Over the past years, several atmosphere models have been developed that account for stellar irradiation and intrinsic heat flux, including atmospheric characteristics such as grains, hazes, and clouds \citep[e.g.][]{Guillot2010, Heng2012, Baudino2017, Molliere2015, Malik2019}. \\
Few earlier works have investigated the impact of the atmospheric conditions including grains and clouds on the planets' long-term thermal evolution: For example, \citet{Kurosaki2017} show that condensation in heavily enriched atmospheres of ice giants accelerates the cooling as the planet emits more energy due to latent heat release. In contrast, \citet{Vazan2013} show a delayed cooling due to an atmosphere enriched in grains for giant planets. 
For isolated, non-irradiated low-mass planets ($M_\mathrm{P}<0.6\,M_J$), \citet{Linder2018} find that including clouds or using different atmospheric codes have only a limited influence on the evolution tracks. \\
In many atmospheric models, the stellar irradiation flux is considered to be the main driver of the physics of the upper atmosphere. ~\citet{Fortney2020} emphasise that not only the equilibrium temperature ($T_{\mathrm{eq}}$) as a measure for the stellar irradiation characterises the atmosphere of giant planets - but also the heat flux from the deep interior (characterised by the intrinsic temperature $T_{\mathrm{int}}$), stressing that the appearance of clouds might as well depend on the atmospheric pressure-temperature  conditions ($P$-$T$) in the deep atmosphere. \\
In coupled atmosphere-interior models with convective interiors, the atmosphere connects to the inner envelope at the radiative-convective boundary (RCB)~\citep[e.g.][]{Thorngren2019, Thorngren2019a, Poser2019}. During the planets' long-term evolution, the RCB moves towards higher pressures for progressing time steps as the planet cools down.
To the best of our knowledge, the effect of deep-seated cloud decks on the RCB and on the long-term thermal evolution of warm irradiated planets has not been studied so far. Knowing the impact on the evolution curves may help characterise giant planets with regard to their heavy element mass and distribution.\\
Therefore, we aim at estimating the influence of atmospheric $P$-$T$ conditions with and without clouds on the long-term evolution of irradiated giant planets. In detail, we vary the atmospheric model while performing thermal evolution calculations for two young warm giant planets, TOI-1268b and WASP-10b. We concentrate our study on warm Jovian planets as they do not show an inflated radius. Inflated gas giants with $T_{\mathrm{eq}}>1000\,$K need an additional amount of extra energy that would induce another model uncertainty~\citep{Thorngren2018, Sarkis+2020}. Also, it is more probable that clouds occur at lower $T_{\mathrm{eq}}$~\citep[e.g.][]{Fortney2020}.\\
In this paper, we look at the difference of cloud-free and cloudy $P$-$T$ structures, exploring how a possibly variable cloud deck location affects the evolution. Our idea is that during evolution, the deep parts of the atmosphere cross the condensation curves of possible cloud condensation curves at different pressure levels, resulting in changing (evolving) cloud deck bases. For our purpose, we apply the approach of~\citet{Poser2019}, adapting and extending the cloud model by~\citet{Heng2012}. In that approach, the semi-grey atmosphere model is employed for coupled interior and thermal evolution calculations. It allows us to approximate the complex radiative transfer and microphysics of cloudy atmospheres in a much simplified manner suitable for this study. A cloud deck is added as an additional absorber in the longwave ignoring shortwave scattering. To understand the general impact of cloud decks in the interior modelling procedure, we also investigate the influence on the $T_{\mathrm{int}}$-$Z_{\mathrm{env}}$ phase space.\\
The paper is structured as follows. Section~\ref{sec:method} outlines our modelling approach to study the thermal evolution of gaseous planets, focusing in Subsections~\ref{sec:atm} and \ref{sec:atm_variability} on the description of the atmospheric model. Subsection~\ref{sec:interior_evolution_model} briefly describes the interior and thermal evolution model, while Subsection~\ref{sec:modelprocedure} summarises the modelling procedure. The planets under study, TOI-1268b and WASP-10b, are introduced in Subsection~\ref{sec:pandsparameters}. Presenting the results, we first show the impact on the \textit{static} $T_{\mathrm{int}}$-$Z_{\mathrm{env}}$ phase spaces for both planets in Section~\ref{sec:interiorspaces}, comparing both cases with and without clouds. \textit{Static} means, our approach matches the planetary mass and radius, without performing additional evolution calculations matching the planets' age. In Section~\ref{sec:radiusevolution_clear}, we show the radius evolution curves for a clear atmosphere. In Section~\ref{sec:radiusevolution_clouds}, we present the main results quantifying how the radius evolution changes for different cloudy models. A discussion of the results follows in Section~\ref{sec:discussion}.

\section{Model}
\label{sec:method}

\begin{figure}
	\includegraphics[width=0.99\columnwidth]{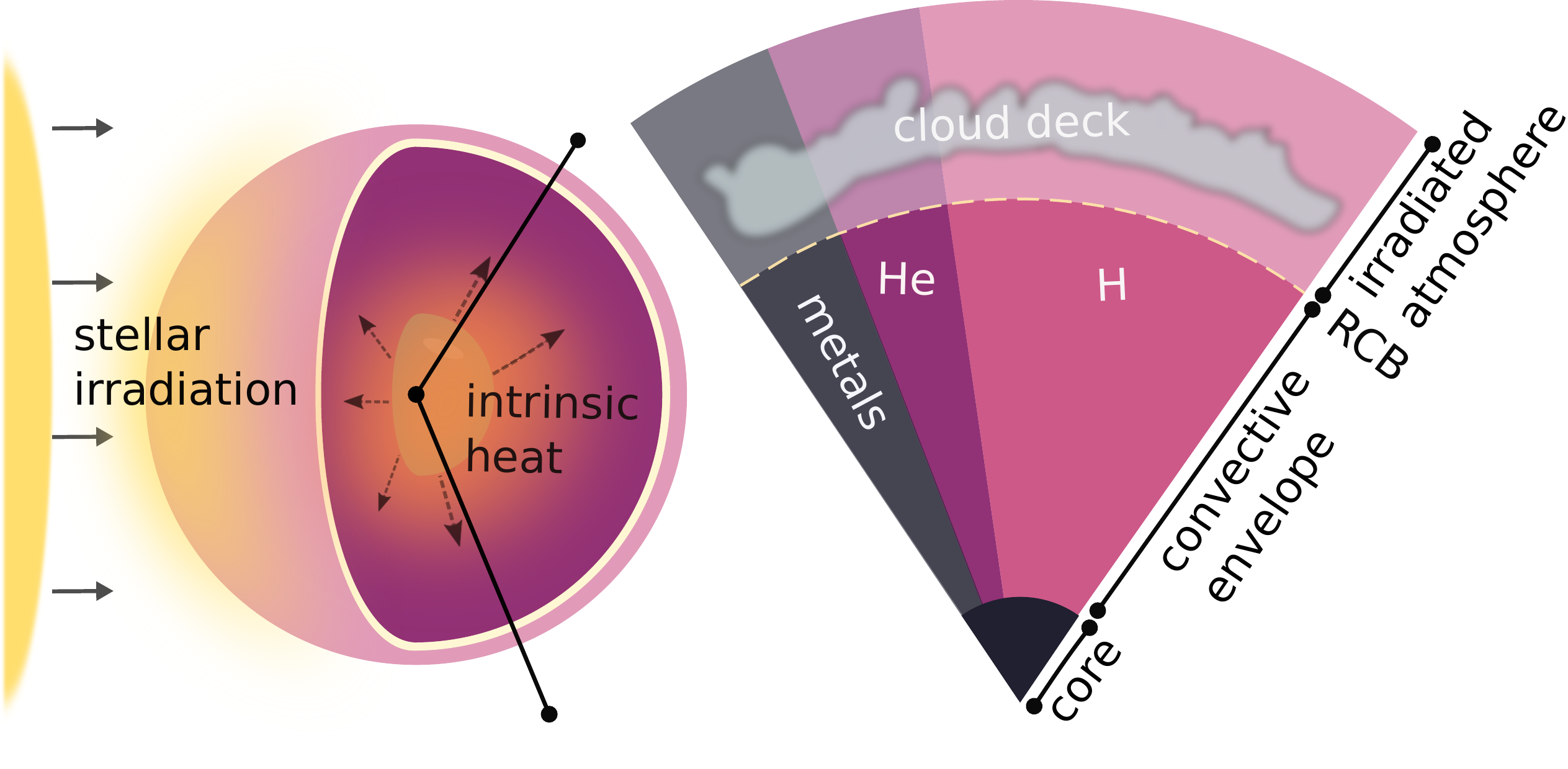}
    \caption{ The model setup in this work: A Jovian planet is irradiated by its star. The atmosphere acts as a bottleneck for the incoming irradiation flux and the outgoing intrinsic heat flux, indicated by arrows. The interior consists of a radiative atmosphere, an outer convective envelope, and an isothermal core. The main constituents are hydrogen and helium, replenished by metals. We consider clouds in the atmosphere and combine atmosphere, interior, and thermal evolution models to obtain the radius evolution of the planet. The atmosphere model is used as an outer boundary condition for the interior model, connecting both models at the radiative-convective boundary. }
    \label{fig:atm_sketch}
\end{figure}

Fig.~\ref{fig:atm_sketch} shows a sketch of the combined atmosphere and interior model setup applied. The planet is made up of up to three layers: a radiative atmosphere, an adiabatic envelope, and an isothermal solid core, following our previous approach in~\citet{Poser2019}. We continue by presenting the atmosphere model accounting for a cloud deck in Subsection~\ref{sec:atm}, followed by a short description of the interior and thermal evolution model in Subsections~\ref{sec:interior_evolution_model} and~\ref{sec:atm_variability}. We outline our modelling process in Subsection~\ref{sec:modelprocedure}, which sets the foundation for the results presented in Section~\ref{sec:results}.

\subsection{Atmosphere}
\label{sec:atm} 

To account for radiative transfer in the atmosphere, we make use of (semi-) analytical 1D, plane-parallel atmosphere models. The clear, cloud-free model goes back to the work of~\citet{Guillot2010}. It depends on the equilibrium temperature $T_{\mathrm{eq}}$, the intrinsic temperature $T_{\mathrm{int}}$, and a semi-grey description of opacities in the short- and longwave wavelength range: $\kappa_\mathrm{S}$, $\kappa_\mathrm{L}$. 
The cloudy atmosphere model is based on the work of~\citet{Heng2012} extending the work of~\citet{Guillot2010} to include the effect of a cloud deck. Here, clouds are added as additional absorbers in the longwave (Section~\ref{sec:subsectioncloudparams}).
These models have been employed in prior studies for coupled atmosphere-interior calculations.~\citep[e.g.][]{Jin2014, Poser2019, Kumar2021, Dietrich2022, MacKenzie2023}.

\subsubsection{The parameter $\gamma$ of the basic grey atmosphere model}
\label{sec:subsectiongamma}
\begin{figure*}
	\includegraphics[width=\textwidth]{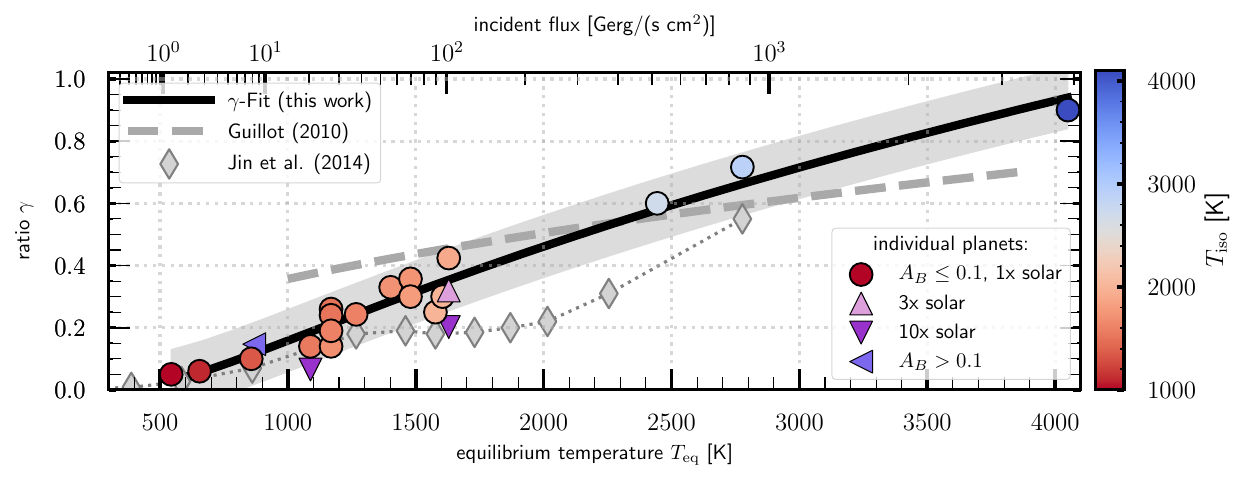}
    \caption{The parameter $\gamma=\frac{\kappa_\mathrm{S}}{\kappa_\mathrm{L}}$ is an essential input parameter for semi-analytical atmosphere models. We derive a $\gamma$-$T_{\mathrm{eq}}$ relation useful for a fast calculation of the temperature-pressure conditions. The resulting fit is shown in black. For comparison, we show the relation by \citet{Guillot2010} in grey dashed and the results from~\citet{Jin2014} in grey diamonds. $T_\mathrm{iso}$ is the temperature of the deep isotherm of the atmospheric $P$-$T$ profile which may extend to several hundred bars. We match the isotherm obtained with the analytical, clear~\citet{Guillot2010} model to the isotherm of previously published atmosphere models, which are based on a more complex solution of radiative transfer and treatment of opacities. Generally, with higher metallicities, the $\gamma$ value shifts to smaller values as the deep isotherm becomes hotter.}
    \label{fig:fitformula_na}
\end{figure*}

Both clear and cloudy atmosphere models account for radiative transfer via semi-grey opacities $\kappa_\mathrm{L}$, $\kappa_\mathrm{S}$. The parameter $\gamma=\kappa_\mathrm{S}/\kappa_\mathrm{L}$ is an essential input to the models, determining the amount of absorption of the incoming flux. To determine the ratio $\gamma$, and consequently to use it for the atmosphere model of our planets, our objective is to find a correlation between $\gamma$ and the equilibrium temperature $T_{\mathrm{eq}}$ of irradiated gas planets.\\
We fitted the $T(P)$ relation by~\citet{Guillot2010} for clear atmospheres to published $P$-$T$ profiles of warm to ultra-hot Jupiters with different equilibrium temperatures by matching the deep isothermal regions manually. The deep isothermal region is characterised by the temperature $T_\mathrm{iso}$. Note, that varying $\kappa_{\mathrm{L}}$ for constant $\gamma$ does not effect the location of the deep isothermal region as the optical depth $\tau$  is proportional to $\propto \kappa_{\mathrm{L}}P$ (assuming constant gravity in the thin atmosphere). But with 
larger $\kappa_{\mathrm{L}}$, the isotherm expands to lower pressures, resulting in a vertical shift of the isotherm. \\
We find the following fit for $\gamma(T_{\mathrm{eq}})$ for ${T_{\mathrm{eq}}=(500-4000)}$K: 
\begin{equation}
\gamma(T_{\mathrm{eq}})=6.24-4.78\,\log (T_{\mathrm{eq}}) +0.92\,(\log(T_{\mathrm{eq}}))^2\,  \mathrm{,}
\label{eq:fitformula}
\end{equation}
which is shown in Fig.~\ref{fig:fitformula_na} in black solid.  We estimate a deviation in $\gamma$ of $\approx \pm 0.1$, based on different possible $P$-$T$-profiles resulting in different isothermal regions. For example, for HD 189733b with $T_{\mathrm{eq}}=1170\,$K, we fit to four different published profiles and get $\gamma=0.14-0.19$.
In Table~\ref{tab:fitformula_gamma}, we present the observational data and publications of the planets used for the fits. 
The equilibrium temperature of each planet is either given in publications or we calculate $T_{\mathrm{eq}} = (1-A)^{1/4}\cdot T_{\mathrm{eq,0}}$ with $ T_{\mathrm{eq,0}}$ as the given zero-albedo equilibrium temperature, and given or estimated albedo $A\approx[0-0.1]$, or via the definition of the irradiation temperature:
The equilibrium temperature is defined as $T_{\mathrm{eq}}=\left((1-A) \cdot f\right)^{1/4}\cdot T_{\mathrm{irr}}$, where the irradiation temperature is $T_{\mathrm{irr}}=T_{\star}\cdot (R_{\star}/a) ^{1/2}$. Here, we set $f=1/4$ as the heat redistribution factor. Other parameters are albedo $A$, $T_\star$ and $R_\star$ are the stellar effective temperature and stellar radius, respectively, and $a$ is the spatial separation between star and planet.
Additionally, we plot planets with higher metallicity or higher albedo in Fig.~\ref{fig:fitformula_na}, see Table~\ref{tab:fitformula_gamma_unused}.  \\
Our resulting fit formula differs from the proposed $\gamma$-relation by~\citet{Guillot2010}, and from the results by \citet{Jin2014}. \citet{Guillot2010} set $\kappa_{\mathrm{L}}=10^{-2}\,$ cm$^2$g$^{-1}$ and $\kappa_{\mathrm{S}}=6\cdot10^{-3}\sqrt{(T_{\mathrm{irr}}/2000\,\mathrm{K})}\,$ cm$^2$g$^{-1}$, specifically adapting these values for HD~209458b.\\
We expect the fit, Eq.~\ref{eq:fitformula}, to be a useful tool when using the \citet{Guillot2010} or \citet{Heng2012} models.  

%%%%%%%%%%%%%%%%%%%%%%%%%%%%%%%%%%%%%%%%%%%%%%%%%%%%%%%%%%%%%%%%%%%%%%%%%%%%%%%%%
\subsubsection{Cloud parameter for the semi-analytical Heng model}
\label{sec:subsectioncloudparams}

\begin{figure}
	\includegraphics[width=0.99\columnwidth]{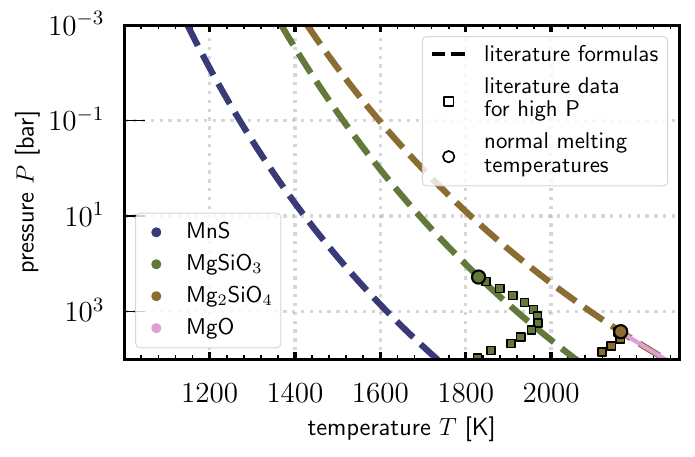}
    \caption{Fits to condensation curves of manganese sulfide (\ch{MnS}), enstatite (\ch{MgSiO3}) and forsterite (\ch{Mg2SiO4}) as used in this work based on previously published  formulas of \citet{Visscher2006, Visscher2010} (dashed) for solar metallicity [M/H]=0. Additionally, we show the values from \citet{Visscher2010} at high pressures (squares) for \ch{Mg2SiO4} and \ch{MgSiO3} as well as their melting temperatures (circles). Their condensation temperatures are depressed for high pressures which is not captured by the fits (dashed).}
    \label{fig:fitformula_condcurve}
\end{figure}

\cite{Heng2012} use a parameterised depiction of a purely absorbing cloud deck where the long-wave opacity $\kappa_\mathrm{L}$ is modified by an additional cloud deck opacity $\kappa_\mathrm{c}$. 
A cloud deck can be added through an additional contribution to the~longwave~opacity~$\kappa_\mathrm{L}$: 
\begin{equation}
\kappa_\mathrm{L}(P)=\kappa_\mathrm{L,0}(P) + \kappa_\mathrm{c}(P)\,\mathrm{.}
\label{eq:longwaveopacity}
\end{equation}
For the cloud-free Guillot model, it is $\kappa_\mathrm{L}=\kappa_\mathrm{L,0}$. For both planets, we use $\kappa_\mathrm{L,0}=10^{-2}\,$ cm$^2$g$^{-1}$ which we found to be an appropriate \textit{mean} Rosseland mean opacity in~\citet{Poser2019}.
The advantage of the Heng model lies in its simplicity of formulation and the reduced number of free parameters. It only allows for the case of a purely absorbing cloud deck without scattering, so that the resulting cloud decks have an warming effect (for $\gamma<1$) - which serves as an upper limit of the influence on the radius evolution. \\
The cloud opacity takes on a Gaussian form, describing an non-uniform cloud deck:
\begin{equation}
\kappa_\mathrm{c}\left(P\right) = \kappa_\mathrm{c,0}\cdot \exp\left[-\Delta_\mathrm{c} \left(1-\frac{P}{P_\mathrm{c}}\right)^2\right] \,  \mathrm{.}
\label{eq:cloudopacity}
\end{equation}
The cloud deck thickness $\Delta_\mathrm{c}$, the cloud deck position $P_\mathrm{c}$ and the cloud opacity normalization $\kappa_\mathrm{c,0}$ account now for an additional opacity in the longwave. A thinner cloud deck corresponds to a larger value of $\Delta_\mathrm{c}$. Please note, that Eq.~\ref{eq:cloudopacity} can easily be extended to several cloud decks (by adding up each decks' $\kappa_\mathrm{c}$), as one expects several cloud decks to be present in an atmosphere, e.g., in Jupiter, Uranus and Neptune~\citep[e.g.][]{West2017, Bjoraker2018, Wong2023, Bhattacharya2023}. However, we decided to use only one cloud deck to minimise the amount of free parameters.
In the following paragraphs, we describe how we choose the free parameter of Eq.~\ref{eq:cloudopacity}.

\ \\
\textit{Cloud deck location $P_\mathrm{c}$}\\
For the purpose of this work, we implicitly assume that clouds are formed by equilibrium processes and that we can comment on potential cloud layers by comparing the clear $P$-$T$ profiles to condensation curves of possible cloud-forming species ~\citep[e.g.][]{Mbarek2016, Ohno2018}. In this work, the intersection between the clear atmospheric profile and the respective condensation curve (e.g. \ch{MgSiO3}) yields the cloud deck pressure $P_\mathrm{c}$. However, the cloud formation process is much more complex and dynamic, as assumed in this work, see \citet{Helling2019, Helling2021a} for reviews. For our work, we need to deduce cloud parameters from studies that include advanced condensation chemistry. \citet{Helling2021} compare cloud properties for hot to ultra-hot gas giants. For the coolest planet in their sample, WASP-43b with $T_\mathrm{eq}=1400\,$K, they find metal oxides (e.g., \ch{SiO}, \ch{MgO}), high temperature condensates (e.g., iron \ch{Fe}) and silicates (e.g., enstatite \ch{MgSiO3} and forsterite \ch{Mg2SiO4}) to be possible condensates, looking at gas pressures $P<10^2\,$bar. However, input to their models are 3D $P$-$T$ structures obtained by global circulation models (GCMs) with a specific (and constant) $T_\mathrm{int}$ value. Our work assumes a deep isotherm extending to a few kbar for low $T_\mathrm{int}$ values for evolved planets which may change pressure range considered for condensation to occur. For our calculations here, we continue by using the condensation curves of \ch{MgSiO3} for WASP-10b and TOI-1268b, when we calculate the intersection \textit{dynamically}, see Subsection~\ref{sec:atm_variability}.\\
Here, we aim at modelling the effect of a (one) possible cloud deck in the deep atmosphere within the coupled thermal evolution calculations. The cloud deck location is the intersection of the pre-calculated clear $P$-$T$ profile with a condensation curve. We use the previously published $P$-$T$ relations for condensation curves by~\citet{Visscher2006, Visscher2010} that are also dependent on the metallicity [M/H]. We show their models for [M/H]=0 ($1\times\,$solar metallicity) in dashed in Fig.~\ref{fig:fitformula_condcurve}. Furthermore, for \ch{Mg2SiO4}, and \ch{MgSiO3} we plot the normal melting temperatures (circled) and the condensation curves for higher pressures (squared) as published by \citet{Visscher2010}. 
Both \ch{Mg2SiO4} and \ch{MgSiO3} are reduced at higher pressures because \ch{SiO} gets replaced by \ch{SiH4}~\citep{Visscher2010}. This behaviour is not included in our fit. Furthermore, \ch{MgO} condenses at $P>10^{3.5}\,$bar~\citep{Visscher2010} which in turn is represented by our fit for \ch{Mg2SiO4} for $P>10^{3.5}\,$bar. In support of this, \citet{Helling2021} study cloud properties for a range of gas giants, e.g. suggesting that metal oxides (SiO, MgO) become more common than silicates (\ch{Mg2SiO4}, \ch{MgSiO3}) at higher pressures.

\begin{figure*}
	\includegraphics[width=0.99\textwidth]{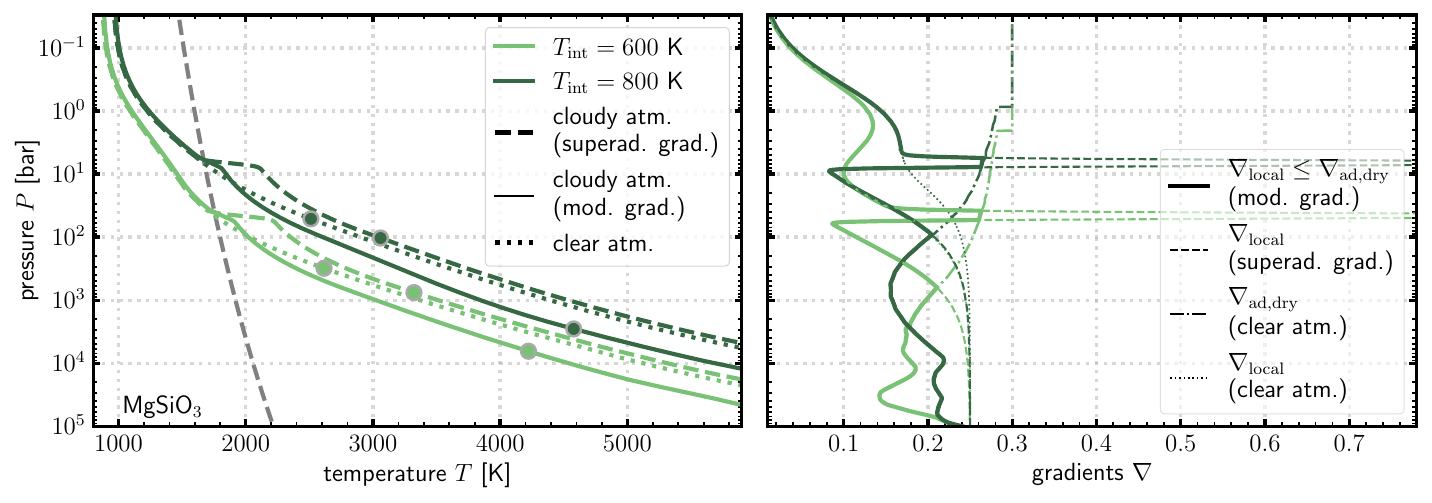}
    \caption{Left: Pressure-temperature profiles of the atmosphere for two intrinsic temperatures $T_{\mathrm{int}}=600, 800\,$K (colour-coded) for a warm Jupiter. The clear $P$-$T$ profile is shown (dotted), as well as the cloudy atmosphere profile \citet{Heng2012} (dashed), and the cloudy $P$-$T$ profile with modified \textit{adiabatic} gradient (solid). Transitions to the interior are marked by a dot. The cloud base $P_\mathrm{c}$ is chosen by the intersection with the condensation curve of $\mathrm{MgSiO_3}$.  Right: The local gradient of the cloudy atmosphere (dashed), the adiabatic gradient of the clear atmosphere (dashed dotted), and the modified gradient of the cloudy atmosphere (solid) are shown. }
    \label{fig:theory_gradients}
\end{figure*}

\ \\
\textit{Cloud deck opacity and thickness}\\
To account for the remaining parameters, the cloud deck opacity $\kappa_\mathrm{c,0}$ and the cloud deck thickness $\Delta_\mathrm{c}$, we compare our model with the $P$-$T$ structure solutions with clouds from~\citet{Linder2018}, in particular the ones simulated with \textsc{PetitCode}~\citep{Baudino2017, Molliere2015, Molliere2017}, and with the cloud opacities of \citet{Helling2014}, \cite{Lee2017}, and \cite{Dobbs-Dixon2013}.
In this study, we use $\kappa_\mathrm{c,0}= 0.05-1.0$ cm$^2$g$^{-1}$ and a cloud deck thickness of $\Delta_\mathrm{c}=[10, 25, 50, 75, 100]$ which we base on the comparison with the results in the papers mentioned above. Following Eq.~\ref{eq:longwaveopacity}, the opacity at cloud level is characterized by a Rosseland mean opacity. Please note that - depending on the cloud location - a Planck mean opacity might be a better approximation, as the Rosseland mean opacity is better suited for regions, as it is the deep atmosphere because it weighs more for wavelengths that contribute a low opacity~\citep{MacKenzie2023, Freedman2014}.

\ \\
\textit{Atmospheric pressure-temperature gradients}\\
\citet{Kurosaki2017} study the effect of low-temperature condensation clouds on the radius evolution using the pseudo-moist adiabatic temperature gradient for the troposphere. Motivated by our previous study and \citet{Kurosaki2017}, we introduce a modification of the local atmospheric gradient, so that it does not exceed the adiabatic gradient of the dry (clear) atmosphere:
\begin{equation}
    \nabla_{\mathrm{local}} \leq \nabla_{\mathrm{ad, dry}} \mathrm{~.}
\end{equation}
Fig.~\ref{fig:theory_gradients} (left) depicts the implications for the $P$-$T$ profile. Fig.~\ref{fig:theory_gradients} (right) shows both the superadiabatic/non-modified (dashed) and the modified gradient (solid) over the same pressure range. Here, for two $T_{\mathrm{int}}$ values, we show the clear profile (dotted) and two cloudy profiles with superadiabatic (dashed) and the modified gradient (solid).\\
First, the warming effect of the cloud deck with the modified gradient is reduced compared to the unmodified gradient.
Furthermore, the $P$-$T$ profiles with the modified gradient result in a cooler adiabat than for the clear case (for a constant $T_{\mathrm{int}}$ of the atmosphere model), and the upper atmosphere is not influenced. This is opposite to \citet{Kurosaki2017}, where the inclusion of condensation shifts the $P$-$T$ profile to lower temperatures for evolving times/cooling of the planet. \\
Further, \citet{Kurosaki2017} connect the atmosphere to the interior at a fixed pressure $P_\mathrm{ad}$ where the (convective) interior starts. In this work, the radiative-convective boundary (RCB) is calculated by Eq.~\ref{eq:RCB}, comparing adiabatic and local gradients of the $P$-$T$ profile of the atmosphere. This yields an individual pair of $(T_{\mathrm{ad}}, P_{\mathrm{ad}})$ for different $P$-$T$ conditions. In Fig.~\ref{fig:theory_gradients}, we plot the RCB as circles. The different $P$-$T$ profiles lead to different $(T_{\mathrm{ad}}, P_{\mathrm{ad}})$ values for the RCB. \\
Despite the differences, we continue with our modification of the atmospheric model description as we expect the reradiated energy (due to latent heat release) does \textit{not} directly influence the adiabat of the interior but the overall (enhanced) emission is then mirrored by the colder adiabat for a given $T_{\mathrm{int}}$ value.

%%%%%%%%%%%%%%%%%%%%%%%%%%%%%%%%%%%%%%%%%%%%%%%%%%%%%%%%
%%%%%%%%%%%%%%%%%%%%%%%%%%%%%%%%%%%%%%%%%%%%%%%%%%%%%%%%
\subsection{Interior and thermal evolution model}
\label{sec:interior_evolution_model}
The interior model is composed of three discrete layers (atmosphere, envelope, core), see Fig.~\ref{fig:atm_sketch}. While the atmosphere and the envelope differ in the assumed energy transport, the mass fractions of hydrogen~$X$, helium~$Y$, and metals~$Z$ are the same in atmosphere and envelope ($X+Y+Z=1$). The helium/hydrogen mass fraction abundance for all models is $Y'\coloneqq M_\mathrm{H}/(M_\mathrm{H}+M_\mathrm{He})=0.27$~\citep{Bahcall1995}. \\
The transition from the radiative atmosphere to the convective interior is determined by the adiabatic and local numeric temperature gradients:
\begin{equation}
\nabla_{\mathrm{local}}\geq \nabla_{\mathrm{ad}} \mathrm{,}
\label{eq:RCB}
\end{equation}
where $\nabla_{\mathrm{ad}}=\left( \frac{\partial\ln{T}}{\partial\ln{P}} \right)_\mathrm{s}$ is taken from the EoS tables.
The radiative-convective boundary (RCB) is then defined by $(T_{\mathrm{ad}}, P_{\mathrm{ad}})$ characterizing the entropy $s$ of the interior adiabat. 
To account for the thermal evolution of the planet, we assume:
\begin{equation}
L_{\mathrm{eff}}-L_{\mathrm{eq}}=L_{\mathrm{int}}=L_{\mathrm{secular}}+L_{\mathrm{radio}} \,\mathrm{.}
\end{equation}

The luminosity $L_{\mathrm{eff}}$ describes the effective luminosity reradiated over the entire surface of the planet. $L_{\mathrm{eq}}=4\pi R_\mathrm{P}^2\sigma_B T_{\mathrm{eq}}^4$ is the absorbed and reemitted stellar flux. The interior heat loss has different contributions:
\begin{equation}
    L_{\mathrm{secular}}=-\int_0^{M_\mathrm{P}} \mathrm{d}m\,T(m,t) \frac{\mathrm{d}s(m,t)}{\mathrm{d}t}
    \label{eq:secularcooling}
\end{equation} 
accounts for cooling and contraction of the planet, and the radiogenic heating is denoted as $L_{\mathrm{radio}}$. For further information on the evolution and interior calculations, see~\citet{Poser2019}. \\
Several other works have shown that the thermal evolution of a planet is influenced by the choice of the equation of state (EoS) for hydrogen, helium, and metals~\citep[e.g.][]{Vazan2013, Miguel2016}. 
We compare the influence of hydrogen and helium EoS by~\citet{Chabrier2021} (CD21) and~\citet{SaumonChabrier1995} (SCvH95) on the clear radius evolution and the static $T_{\mathrm{int}}$-$Z_{\mathrm{env}}$ phase space in Subsection~\ref{sec:interiorspaces}. For the cloudy radius evolution, we use the newer H/He - EoS by CD21 (Subsection~\ref{sec:radiusevolution_clouds}). Note that the SCvH95 - EoS is based on a chemical model, while that of CD21 includes ab initio data in particular for the warm dense matter region. The metals of the envelope ($Z_{\mathrm{env}}$) are represented as ice, while the core is made of rocks (both EoS from~\cite{Hubbard1989a}). \\
Other model assumptions that may influence the heat transport and consequently the thermal evolution of the planet are the possibility of thermal boundary layers~\citep[e.g.][]{Nettelmann2016, Scheibe2019, Scheibe2021, Bailey2021}, inefficient convection due to compositional gradients causing double-diffusive or layered convection~\citep[e.g.][]{Stevenson1985,Leconte2012, Leconte2013}, or non-adiabatic interiors ~\citep{Debras2019, Debras2021}.

\subsection{Variable cloud deck location during the long-term evolution}
\label{sec:atm_variability}
\begin{figure*}
 	\includegraphics[width=\textwidth]{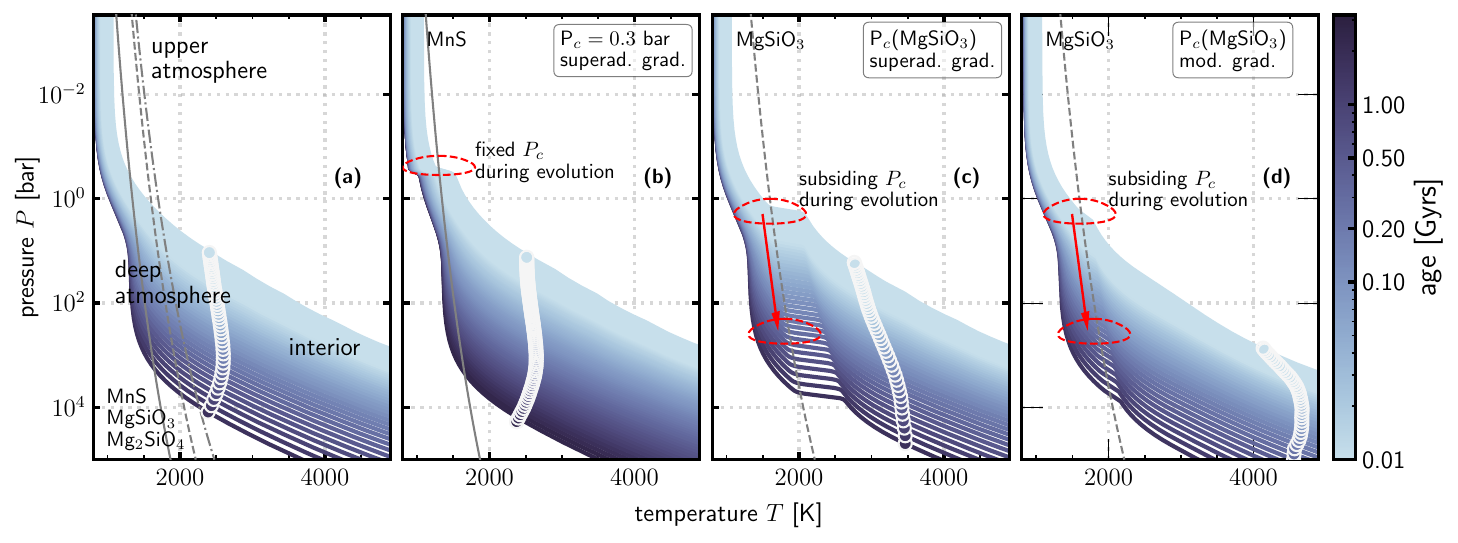}
    \caption{Pressure-temperature profiles of the atmosphere for WASP-10b under clear (a) and cloudy (b-d) conditions for $T_{\mathrm{int}}=200-1100$ K. The condensation curves ([M/H=0]) of different cloud-forming species (grey) cross the entire atmosphere during the $T_{\mathrm{int}}$ evolution. The radiative-convective boundary (dot) connects the atmosphere to the interior. The clear $P$-$T$ atmosphere model (a) serves as a starting point for the cloud deck location of the cloudy atmosphere models. The three right subfigures show the resulting $P$-$T$ conditions during evolution with a fixed cloud deck location ($P_\mathrm{c}=0.3\,$bar) (b), and a \textit{dynamic} cloud deck with subsiding $P_\mathrm{c}$ (c, d). Parameters are $\kappa_\mathrm{c,0}=0.1\,$cm$^2$g$^{-1}$, $\Delta_\mathrm{c}=75$. The profiles are coloured based on the age of the planet, determined through thermal evolution calculations. The corresponding radius evolution curves are shown in the second row of Fig.~\ref{fig:WASP10b_evol_overview}. }
    \label{fig:WASP10b_PT}
\end{figure*}

 In this work, we want to investigate the effect of the cloud deck location during the planets' thermal evolution with two approaches. The first uses a \textit{fixed} cloud deck location for all $P$-$T$ profiles for the various $T_{\mathrm{int}}$ values during the evolution of the planet. The second, which we denote further as \textit{dynamic} or \textit{subsiding $P_\mathrm{c}$}, uses the intersection between the pre-calculated clear profile (for each $T_{\mathrm{int}}$ value) and the respective condensation curve. With that approach, the effect of cloud decks deep in the atmosphere at several hundred bar can be modelled at low $T_{\mathrm{int}}$ values (as for an old or already cooled down planet). \\
 We show the effects of both approaches on the $P$-$T$ profiles in Fig.~\ref{fig:WASP10b_PT}, using WASP-10b as an example. Starting on the left, Fig.~\ref{fig:WASP10b_PT} (a) shows clear, non-cloudy, $P$-$T$ profiles for $T_\mathrm{int}=300-1000\,$K. They intersect the condensation curves of \ch{MnS}, \ch{MgSiO3}, and \ch{Mg2SiO4} at high pressures ($P>1\,$bar). Then, we show the $P$-$T$ profiles obtained with a fixed cloud deck pressure $P_\mathrm{c}=0.3\,$bar in subfigure (b). The very right subfigures (c) and (d) show the results obtained with the \textit{dynamic} approach where $P_\mathrm{c}$ is subsiding with the evolving age of the planet. 
 With that, we mimic an atmosphere variable in time - not only in $T_{\mathrm{int}}$, but also in the total cloud opacity $\kappa_\mathrm{c}(P, P_\mathrm{c})$, see Eq.~\ref{eq:cloudopacity}. \\
Specifically, for WASP-10b and TOI-1268b, we take the intersections with \ch{MgSiO3} when we calculate the intersection \textit{dynamically}. Else, for the \textit{fixed} case, where the cloud base pressure $P_\mathrm{c}$ is constant throughout the thermal evolution, we choose $P_\mathrm{c}=1, 10$ bar for TOI-1268b and $P_\mathrm{c}=0.3$ bar for WASP-10b.

\subsection{Overview of the model procedure}
\label{sec:modelprocedure}

In this Subsection, we give an overview of our method. The key components are as follows:

\begin{enumerate}
    \item \textbf{Interior structure} We construct the planetary profile along the mass coordinate $m$ —  mass $m$, radius $r$, temperature $T$, pressure $P $, density $\rho$, and entropy $s$ — from $m=0$ to $m=M_\mathrm{P}$. This model matches the given (observationally derived) planetary mass $M_\mathrm{P}$ and radius $R_\mathrm{P}$. Assuming a bulk composition of hydrogen, helium, and metals with a fixed H/He ratio and a specified metal mass fraction for the atmosphere $Z_\mathrm{atm}$ and envelope $Z_\mathrm{env}$, our model iteratively determines the core mass $M_{\mathrm{core}}$ to obtain $M_\mathrm{P}$: 
    \begin{align}
    M_\mathrm{P} &= M_\mathrm{atm}+M_\mathrm{env} + M_\mathrm{core}\quad  \mathrm{,}
    \end{align}
   where $M_{\mathrm{atm}}$ and $M_{\mathrm{env}}$ are the total mass of the atmosphere and envelope of the three layer model.\\ 
   \textbf{Static $T_{\mathrm{int}}$-$Z_{\mathrm{env}}$ phase space} 
     The intrinsic temperature $T_{\mathrm{int}}$ of the planet is a crucial input, dictating the internal heat. A higher $T_{\mathrm{int}}$ results in a lower envelope density and thus variations in $M_{\mathrm{Z,env}} $ for a given $Z_\mathrm{env}$, affecting the core mass $M_{\mathrm{core}}$ to align with the given planetary mass. For $Z_\mathrm{env}=Z_\mathrm{atm}$, it is for the total planetary metallicity $Z_\mathrm{P}$ with the total mass of the metals $M_\mathrm{Z}$:
    \begin{align}
     Z_\mathrm{P}&=M_\mathrm{Z}/M_\mathrm{P}  =Z_\mathrm{env}\,(M_\mathrm{atm}+ M_\mathrm{env})/M_\mathrm{P}+M_\mathrm{core}/M_\mathrm{P}  \quad \mathrm{.}
    \end{align}
    The $T_{\mathrm{int}}$-$Z_{\mathrm{env}}$-$Z_\mathrm{P}$ phase space is explored in Subsection~\ref{sec:interiorspaces} to understand the relationship between these parameters.
    Higher $T_{\mathrm{int}}$ values typically lead to higher values of $Z_\mathrm{P}$ due to an increasing core mass, as depicted in Fig.~\ref{fig:TOI1268b_ZPTint}. Here, the x-axis label, \textit{present intrinsic temperature} $T_\mathrm{int}$, is intended to illustrate that the models displayed are possible solutions that yield the observed $M_\mathrm{P}$ and $R_\mathrm{P}$ (but not necessarily the age constraint).
    In order to make Fig.~\ref{fig:TOI1268b_ZPTint}, we calculate for $Z_\mathrm{env}=0-0.52$ (in intervals of $0.01$) for $T_\mathrm{int}=0-1000$ K (in intervals $10$ or $50$ K) single interior structure models. That makes for each bulge $\approx 1000-5000$ models. We do not employ any Bayesian or Markov chain Monte Carlo (MCMC) methods in our algorithm, as it has not been specifically designed to accommodate such techniques at this time. We refer to this as \textit{static}, indicating that the phase space illustrates the influence of the model assumptions and does not consider the thermal evolution of the planet but instead represents the parameter space.
    
    \item \textbf{Radius evolution} 
    By considering the age of the planetary system, we set tighter constraints on the metal content of the planet. First, for a fixed set of $M_\mathrm{P}$, $M_\mathrm{core}$, $Z_\mathrm{env}$  we calculate interior models for $\approx 60-80$ $T_\mathrm{int}$ values in the range of $100-1000\,$K. Here, the free parameter is the radius of the planet $R_\mathrm{P}$: the higher $T_\mathrm{int}$, the larger $R_\mathrm{P}$. Applying Eq.~\ref{eq:secularcooling} allows us to calculate the evolution of the radius of the planet, $R_\mathrm{P} (t)$. The results are shown in Subsections~\ref{sec:radiusevolution_clear} and ~\ref{sec:radiusevolution_clouds}.
\end{enumerate}

\section{Results}
\label{sec:results}
We introduce the planets for this study in Section \ref{sec:pandsparameters}. We then continue to present the results for the static $T_{\mathrm{int}}$-$Z_{\mathrm{env}}$ phase space in Section~\ref{sec:interiorspaces}. From a modelling perspective, the $T_{\mathrm{int}}$-$Z_{\mathrm{env}}$ phase space contains information on the metal content of the planet. We present the impact of the observational uncertainties in mass and radius on the $T_{\mathrm{int}}$-$Z_{\mathrm{env}}$ phase space, the impact of the H/He-EoS (SCvH95 vs. CD21), and finally, the impact when including the cloud models. 
While our main focus lies in understanding the impact of clouds on the radius evolution, the radius evolution with a clear atmosphere (no clouds) serves as anchor point for the comparison. Subsequently, in Section~\ref{sec:radiusevolution_clear}, we present the clear radius evolution for both planets. The key results, the effects of various cloud models on the radius evolution, are presented in Section~\ref{sec:radiusevolution_clouds}. 

\subsection{Choice of planets}
\label{sec:pandsparameters}

We consider two planets similar in their young age and equilibrium temperature, but with different densities and masses.
The first one, TOI-1268b \citep{Subjak2022, Dong2022}, has a similar mass as the hot Saturn WASP-39b, but a higher density of $\rho = 0.53\,\rho_{\mathrm{Jup}}$ due to its smaller radius compared to WASP-39b's density of $\rho = 0.14\,\rho_{\mathrm{Jup}}$~\citep{Faedi2011}. With an age of $110$-$1000\,$Myrs, it falls within a group of young ($<1\,\mathrm{Gyr}$) gas giants with measured masses and radii, making the planet a candidate for testing evolution and formation theories. TOI-1268b resides at the inflation threshold~\citep{Sarkis+2020} and is probably not inflated. 
The second planet, WASP-10b, is characterised by an age of $190$-$350\,\mathrm{Myrs}$~\citep{Johnson2009,Christian2009,Maciejewski2011}, a mass of $2.96\,M_{\text{Jup}}$, and a planetary radius of $1.02\,R_{\rm Jup}$~\citep{Maciejewski2011a}. We extend our studies of WASP-10b in~\citet{Poser2019} to this paper. The observational parameters used in this work are listed in Table~\ref{tab:starplanetparam}. \\
We have chosen the planets for different reasons: First, they reside at the inflation limit. This allows us to reduce a possible additional uncertainty due to the amount of extra energy needed to explain the inflated radius~\citep{Thorngren2018, Sarkis+2020}. Second, it is more probable that clouds occur at lower $T_{\mathrm{eq}}$~\citep[e.g.][]{Fortney2020}. However, the planets differ in mass and overall density, comparing possible arising uncertainties for both a warm Saturn (TOI-1268b) and a warm Jupiter (WASP-10b). Furthermore, the investigation of explicit young planets contributes to understanding formation and (early) evolution processes.
\begin{table}
\caption{Stellar and planetary parameters}
\label{tab:starplanetparam}
\centering
\begin{tabular}{c|cc}
\hline 
 & \textbf{TOI-1268b}$^{5}$  & \textbf{WASP-10b}	\\
\hline 
planetary parameter &&\\
$M_\mathrm{P}\,[M_{\text{Jup}}]$           	&$ 0.29\pm0.04$    & $2.96^{+0.22}_{-0.17}$ $^{1}$	 \\
$R_\mathrm{P}\,[R_{\text{Jup}}]$		          &   $0.82\pm0.06$    & $1.03^{+0.077}_{-0.03}$ $^{4}$	  \\
$T_{\text{eq,A=0}}\,[K]$		 	& $919$        & $ 950^{+30}_{-26}\,$	$^{4}$	  \\ 
$\rho\,$[$\rho_{\text{Jup}}$] & $0.53$  & $1.43$\\
\hline
orbital parameter &&\\
e		                          & $0.09^{+0.04}_{-0.03}$      & $0.013\pm0.063$	 $^{3}$	    \\
a [AU]	                            & $0.072\pm0.01$     & $0.0369^{+0.0012}_{-0.0014}\,$	$^{1}$	  \\
P [d]	                           & 8.15           & 3.09	   \\
\hline 
stellar parameter &&\\
$M_{\star}\,[\text{M}_{{\odot}}]$		& $0.96\pm0.04$ & $0.75\,$ $^{2}$	 \\
$R_{\star}\,[\text{R}_{\odot}]$			& $0.92\pm0.06$ & $0.67\,$ $^{4}$	 \\
$T_{\star}\,$ [K]		           		 & $5300\pm100\:$    & $4675\pm100\:$	$^{1}$	 \\
age $\tau_{\star}$ [Myr]		         	 & $110-1000\,$    & $270\pm80\,$ $^{3}$	  \\
\hline 
\end{tabular}
\raggedright{\\
\textbf{References.} $^{1}$~\cite{Christian2009}, $^{2}$~\cite{Johnson2009}, 
$^{3}$~\cite{Maciejewski2011}, $^{4}$~\cite{Maciejewski2011a}, $^{5}$~\cite{Subjak2022}.
}
\end{table}

\subsection{Static $T_{\mathrm{int}}$-$Z_{\mathrm{env}}$ phase spaces}
\label{sec:interiorspaces}

\begin{figure*}
\begin{minipage}{0.99\textwidth}
    \includegraphics[width=0.99\columnwidth]{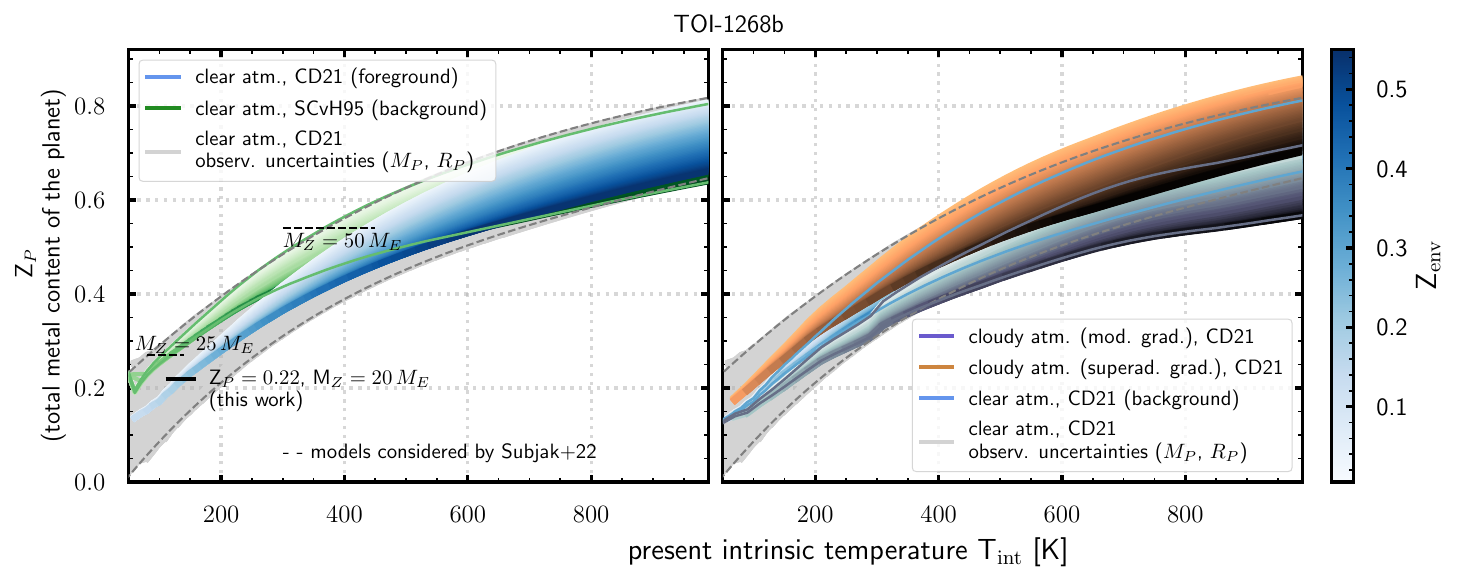}
\end{minipage}
\begin{minipage}{0.99\textwidth}
    \includegraphics[width=0.99\columnwidth]{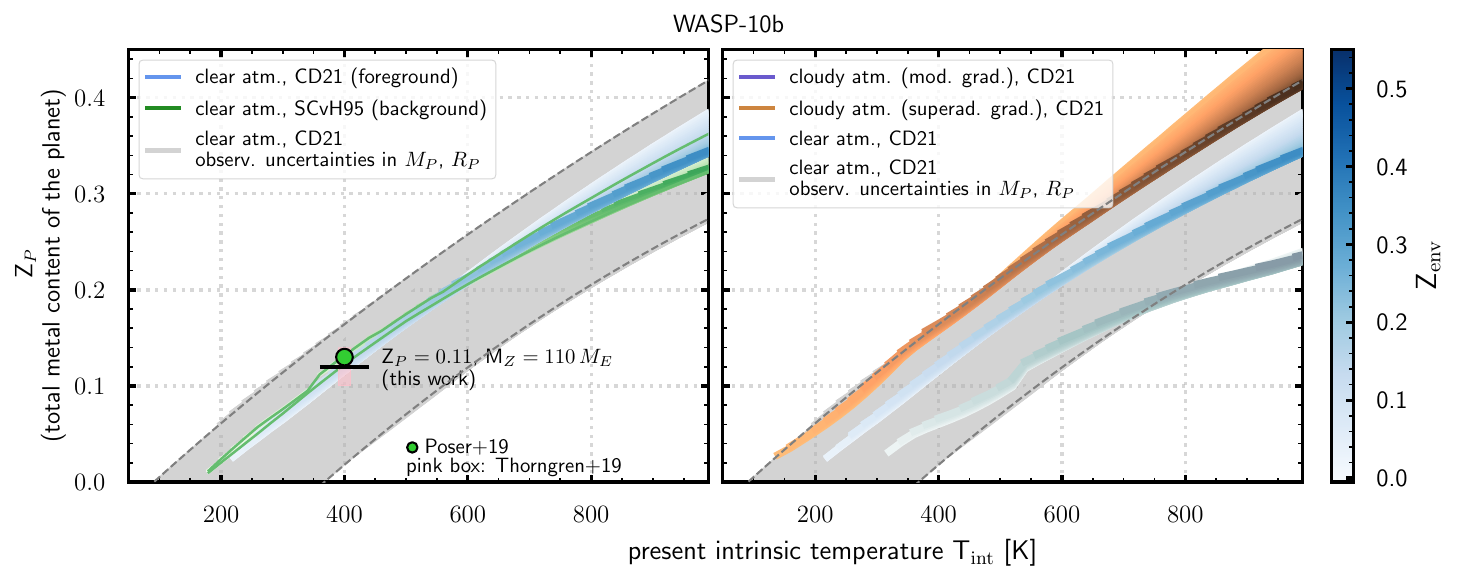}
\end{minipage}
    \caption{Phase space ($T_{\mathrm{int}}-Z_\mathrm{P}-Z_{\mathrm{env}}$) of our interior model setup for TOI-1268b and WASP-10b, highlighting the influence of uncertainties in mass and radius, EoS and atmosphere model. The higher the present $T_{\mathrm{int}}$, the more metals can (theoretically) be included in the envelope (with higher $Z_{\mathrm{env}}$ values possible), and subsequently a higher $Z_\mathrm{P}$ may arise. The same degree of colour shade is used in the upper and lower panels (lighter shade: lower $Z_{\mathrm{env}}$, darker shade: higher $Z_{\mathrm{env}}$). The vertical black lines show $Z_\mathrm{P}$ values obtained by other works: for TOI-1268b, the black dashed line indicates the $Z_\mathrm{P}$ values considered by~\citet{Subjak2022}, whereas the solid lines indicate the derived $Z_\mathrm{P}$ value from this work. The contour lines of models in the background are shown for visualization. Additionally, the grey area, contoured by grey dashed lines, indicates the resulting model space when taking the observational uncertainties in $M_\mathrm{P}$ and $R_\mathrm{P}$ into account (shown for CD21 EoS and a clear atmosphere model).}
    \label{fig:TOI1268b_ZPTint}
\end{figure*}

\subsubsection{General description of the $T_{\mathrm{int}}$-$Z_{\mathrm{env}}$ phase space and approach}
We here show the static $T_{\mathrm{int}}$-$Z_{\mathrm{env}}$ phase space, spanning the ($T_{\mathrm{int}}-Z_\mathrm{P}-Z_{\mathrm{env}}$)-plane. We want to see how the space changes when applying different model atmospheres, comparing to the reference clear atmosphere case. We compare the effects to the observational uncertainties and the impact of different H/He-EoS. \\
The first step towards the static interior modelling is to find the $\gamma$ value of a clear atmosphere using the fit-formula, see Eq.~\ref{eq:fitformula}. We find $\gamma=0.152$  for TOI-1268b ($T_\mathrm{eq}(T_\star, R_\star, \mathrm{a})=913$ K) and $\gamma=0.167$ for WASP-10b ($T_\mathrm{eq}(T_\star, R_\star, \mathrm{a})=960$ K). Interestingly, we find that in the case of WASP-10b, our parameter choice results in the same $P$-$T$ structure as in \cite{Poser2019}, fitted to a known $P$-$T$ structure. We conclude that the fit formula gives a good first-order estimate of the $P$-$T$ structure for irradiated planets with an unknown $P$-$T$ structure. Continuing our previous work on WASP-10b and WASP-39b, we set $\kappa_{\mathrm{L,0}}=0.01\,$cm$^2$g$^{-1}$ (equivalent to $\kappa_{\mathrm{L}}$ for the clear model).\\
In the second step, we model the $T_{\mathrm{int}}$-$Z_{\mathrm{env}}$ phase space: The combination of a given planetary mass and radius leads to an inferred heavy element mass $Z_\mathrm{P}$. Without the age constraint of the system, the parameter space of possible solutions can be large, as shown in Fig.~\ref{fig:TOI1268b_ZPTint}. The parameter space is a function of the atmosphere and interior model setup, and input parameter, e.g., the EoS and distribution of the heavy elements (fully mixed planet versus all metals in the core). A possible constraint for the envelope metallicity $Z_{\mathrm{env}}$ can be given by the derived atmospheric metallicity~\citep[e.g.][]{Wakeford2018,Muller2023a,Poser2019}. 

In Fig.~\ref{fig:TOI1268b_ZPTint} we show the relation of the total heavy element content $Z_\mathrm{P}$, the metal distribution (displayed as $Z_{\mathrm{env}}$), and the internal heat flux, represented by the intrinsic temperature $T_{\mathrm{int}}$ for TOI-1268b (upper row) and WASP-10b (lower row). 
For both planets, the influence of different H/He equation of state and the influence of observational uncertainty are shown in the first column, the possible degeneracy due to the chosen atmospheric model is shown in the second column.  \\
In general, with hotter interiors (higher $T_{\mathrm{int}}$), the more heavy elements can be included in the planet. The uppermost line of each bulge in each subfigure highlights the case in which all metals are in the core ($Z_{\mathrm{env}}=0$). For a fully mixed planet ($M_\mathrm{core}=0$), the maximum $Z_{\mathrm{env}}$ is then the lower limit of the $T_{\mathrm{int}}$-$Z_{\mathrm{env}}$ phase space bulge: For a given $T_{\mathrm{int}}$, the total metal content of the planet is higher if all metals reside in the core than in the envelope (due to different EoS for the metals in the core and envelope). \\
The matching $T_{\mathrm{int}}$ value for the observed mass, radius, and age has to be determined with calculations of the thermal evolution, see Section~\ref{sec:radiusevolution_clear}.

\subsubsection{Impact of the observational uncertainties in planetary mass and radius}
The left subfigures show the $T_{\mathrm{int}}$-$Z_{\mathrm{env}}$ phase space with a clear, non-cloudy atmospheric model. The blue-shaded bulge in the foreground shows the $T_{\mathrm{int}}$-$Z_{\mathrm{env}}$ phase space with the CD21 EoS for H/He. The uni-colour grey area depicts the uncertainty in observational mass and radius, displayed for the model setup using the CD21 EoS for H/He. It is the combination of $M_\mathrm{P}$ and $R_\mathrm{P}$ resulting in the largest and smallest density within the observational uncertainty (e.g., for the most dense combination of WASP-10b: $M_\mathrm{P}=3.18\,M_\mathrm{Jup}$, $R_\mathrm{P}=1.00\,R_\mathrm{Jup}$). The most probable value for $Z_\mathrm{P}$ lies then in between the grey dashed boundaries. A statistical analysis is left for future work. We point out that~\citet{Muller2020} investigated the influence of observational uncertainties on the modelling process. \\
Generally, the $T_{\mathrm{int}}$-$Z_{\mathrm{env}}$ phase space for WASP-10b is narrower than the one for TOI-1268b due to the different bulk densities. For TOI-1268b, it is possible to include up to $Z_\mathrm{P}=0.8$ for high intrinsic temperatures ($T_{\mathrm{int}}=1000\,$K). For the same intrinsic temperature, WASP-10b may include only up to $Z_\mathrm{P}=0.4$.

\subsubsection{Impact of the H/He-EoS}
The green bulges in the background show the results using SCvH95 EoS - instead of the newer CD21 - for H/He. For TOI-1268b, for a given $T_ {\mathrm{int}}<600\,$K, we can include more metals in the planets using SCvH95 EoS instead of CD21 EoS. The newer CD21 EoS leads to denser planets than SCvH95, so that we need a smaller core mass to match $M_\mathrm{P}$, leading to a smaller $Z_\mathrm{P}$. For $T_ {\mathrm{int}}>600\,$ K, using CD21 leads to slightly higher metallicities than using SCvH95.
For WASP-10b, we see a similar behaviour with a turning point at $T_ {\mathrm{int}}\approx550\,$K. For $T_ {\mathrm{int}}<550\,$K, the difference between CD21 and SCvH95 is not as large as for TOI-1268b and increases for higher $T_ {\mathrm{int}}$.\\
A comparison between SCvH95, CD21 and the EoS published by \citet{Chabrier2019} (CMS19) for H/He has been investigated for the evolution tracks of brown dwarfs by~\citet{Chabrier2023}. As they point out, the new CD21 EoS leads to cooler isentropes in the convective interiors than SCvH95 and CMS19. We can see this behaviour as well when we can include more metals (for  $T_ {\mathrm{int}}<600\,$K), see also Subsection~\ref{sec:radiusevolution_clear} for a comparison in the radius evolution. 

\subsubsection{Impact of the atmosphere model including clouds}
The right subfigures show the $T_{\mathrm{int}}$-$Z_{\mathrm{env}}$ phase space using cloudy atmosphere models instead of a clear model. For both planets, we depict the model space for two different sets of cloud models. For \mbox{TOI-1268b}, we use $\kappa_\mathrm{c,0}=0.1\,$cm$^2$g$^{-1}$ and $\Delta_\mathrm{c}=10$, and as cloud base $P_\mathrm{c}$ the intersection with the condensation curve of \ch{MgSiO3} which may differ for each $T_{\mathrm{int}}$ value. For WASP-10b, we use $\kappa_\mathrm{c,0}=0.2\,$cm$^2$g$^{-1}$ and $\Delta_\mathrm{c}=10$, and the same $P_\mathrm{c}$(\ch{MgSiO3}). In both subfigures, the orange model space uses the non-modified (superadiabatic) cloud gradient, while the grey-lilac model space employs the modified gradient for the ($P$-$T$) profile of the atmosphere.\\ 
Including a purely absorbing cloud layer in the atmosphere alters the phase space. Inserting the cloud deck with superadiabatic cloud gradient allows for a higher metal content (higher $Z_\mathrm{P}$) in the envelope compared to the clear case (as shown in \citet{Poser2019}). We can see that as the orange bulge shifts to higher metallicities. This is because the change of the atmosphere model with the non-modified gradient shifts the interior model towards higher entropies which leads to a higher $Z_\mathrm{P}$. 
The case of the cloud deck with modified gradient (grey-lilac model space), conversely, restricts the amount of metals within the planet's interior as the atmosphere model leads to lower entropies compared to the clear atmosphere model. \\
This behaviour is mainly determined by the RCB. The location of the RCB plays an important role for the $T_{\mathrm{int}}$-$Z_{\mathrm{env}}$ phase space, see also \citet{Thorngren2019a}. In our model, it is self-consistently determined by the adiabatic and local gradients in the atmosphere. Compared to the clear model, the cloudy atmosphere model with the modified gradient shifts the RCB to lower pressures and temperatures. As a result, the entropy is shifted to lower values, leading to a colder interior. \\ 
For TOI-1268b, the impact of the two cloud models chosen here, is of about $10\%$ in $Z_\mathrm{P}$. For example, for $T_{\mathrm{int}}=200\,$K, we obtain a span of the total heavy element content of $Z_\mathrm{P}=0.25-0.35$ between both cloud models. \\ 
For the higher density WASP-10b, the narrower form of the clear atmosphere $T_{\mathrm{int}}$-$Z_{\mathrm{env}}$ phase space is mirrored for the cloudy atmosphere. Compared to TOI-1268b, the differences of the cloudy atmosphere spaces to the clear case are larger due to a larger grey cloud opacity and lie in the range of the observational uncertainties.\\
One can see a bend of the cloudy atmosphere model with modified gradient (grey-lila bulge) at $T_\mathrm{int}\approx 500\,$K. This may be due to the calculation of the RCB and the ensuing adiabat. When calculating the RCB, the atmospheric local gradient is compared with the adiabatic gradient which we get from the EoS tables. As soon as the local gradient becomes larger than the adiabatic, we set the RCB, see Eq.~\ref{eq:RCB} and Fig.~\ref{fig:theory_gradients}. For smaller $T_\mathrm{int}$ values, we see that the intersection shifts to higher temperatures, so that we have a bend in the RCB in the $P$-$T$ space, similar to that seen in Fig.~\ref{fig:WASP10b_PT} (d) where the RCB (white circled dots), is in the range of $4000-5000\,$K. 

\subsection{Clear radius evolution}
\label{sec:radiusevolution_clear}

\begin{figure}
\begin{minipage}{0.49\textwidth}
	\includegraphics[width=0.99\columnwidth]{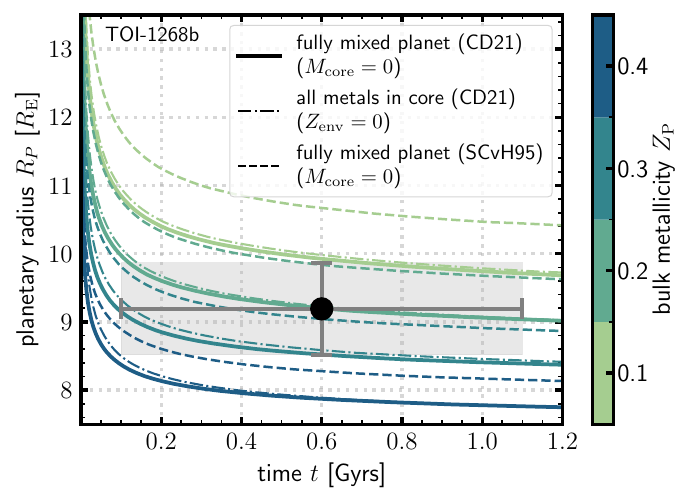}        \end{minipage}
    \begin{minipage}{0.49\textwidth}
    \includegraphics[width=0.99\columnwidth]{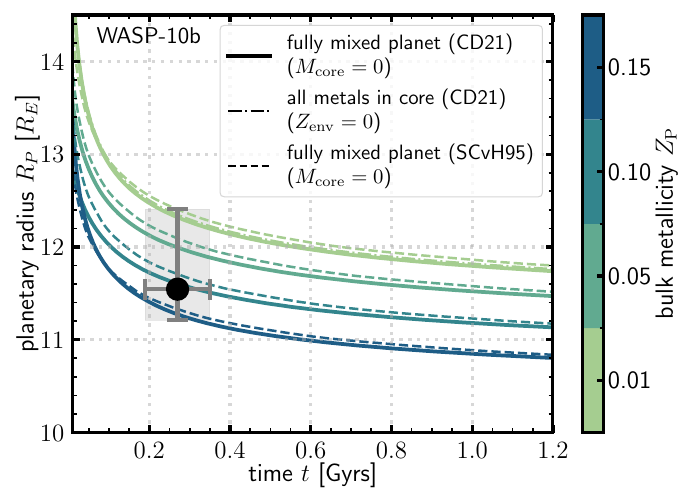}
    \end{minipage}
    \caption{Radius evolution curves for TOI-1268b and WASP-10b with a clear atmosphere. The grey box represents the uncertainty in measured radius and age. For TOI-1268b, we compare the H/He EoS of CD21 (solid) and SCvH95 (dashed) on the radius evolution. Additionally, evolution curves for different total $Z_\mathrm{P}$ in the planet (different colours) are shown, as well as the influence of the metal distribution inside the planet: The planet cools faster for a fully mixed (solid) versus all metals in the core (dash-dotted). For WASP-10b, the difference between a fully mixed planet versus all metals in the core is smaller than for TOI-1268b. Here, a slight difference is noticeable only for $Z_\mathrm{P}=0.01$.}
    \label{fig:theory_evolution}
\end{figure}

In Fig.~\ref{fig:theory_evolution}, we plot several possible evolution curves for TOI-1268b and WASP-10b. For both planets, we show the influence of the H/He-EoS, and in the case of TOI-1268b, we additionally show the effect of the metal distribution. \\
For TOI-1268b, the difference in H/He EoS of CD21 (solid) and SCvH95 (dashed) is up to $1\,R_\mathrm{E}$. It mirrors the results from the static $T_{\mathrm{int}}$-$Z_{\mathrm{env}}$ phase space, where a difference between the EoS is apparent.
The larger $Z_\mathrm{P}$, the smaller the difference between both EoS, as the amount of H/He decreases. \citet{Muller2020} showed the same systematic for the use of CMS19 and SCvH95 EoS.
\\
To derive the metallicity of the planets, \citet{Subjak2022} compares its planetary mass and radius with the isochrones calculated from the evolution curves of~\citet{Fortney2007}, finding a total heavy element mass of $M_\mathrm{Z}=50~M_\mathrm{E}$ to be a good fit. We modelled the thermal evolution within our model setup and found, in contrast to~\citet{Subjak2022}, a heavy element content of about $\approx15-25\,M_\mathrm{E}$ ($Z_\mathrm{P}\approx 0.1-0.3$) that matches the observed mass, radius, and age, see Fig.~\ref{fig:theory_evolution} for TOI-1268b. The best match is for $Z_\mathrm{P}=0.22$ ($M_\mathrm{Z}=20\,M_\mathrm{E}$). This would correspond to the present intrinsic temperatures $\approx160\,$K. The heavy element content would be similar to the expected amount of heavy elements in the cold Solar System analogue Saturn \citep[e.g.][]{Helled2019}. We assume the difference of our findings and those of \citet{Subjak2022} primarily stems from the H/He EoS by~\citet{Fortney2007} (SCvH95).\\
In contrast to TOI-1268b, for the heavier WASP-10b, the difference between both EoS for the radius evolution is more subtle and the derived metal content would not change much. For WASP-10b, we find a heavy element content of $M_\mathrm{Z}\approx0-130\,M_\mathrm{E}$ ($Z_\mathrm{P}\approx0-0.15$) which is consistent with the findings of others, for example \citet{Thorngren2019}. Looking at the static $T_{\mathrm{int}}$-$Z_{\mathrm{env}}$ phase space for WASP-10b, the green point in Fig.~\ref{fig:TOI1268b_ZPTint} shows the interior model used in \cite{Poser2019} whereas the underlying pink box show the probable $T_{\mathrm{int}}$-$Z_{\mathrm{env}}$ phase space suggested by \citet{Thorngren2019}. They used a Bayesian model to infer the metallicity of the planets, placing an upper limit on the atmospheric metallicity.

\subsection{Cloudy radius evolution}
\label{sec:radiusevolution_clouds}

\begin{figure*}
	\includegraphics[width=0.99\textwidth]{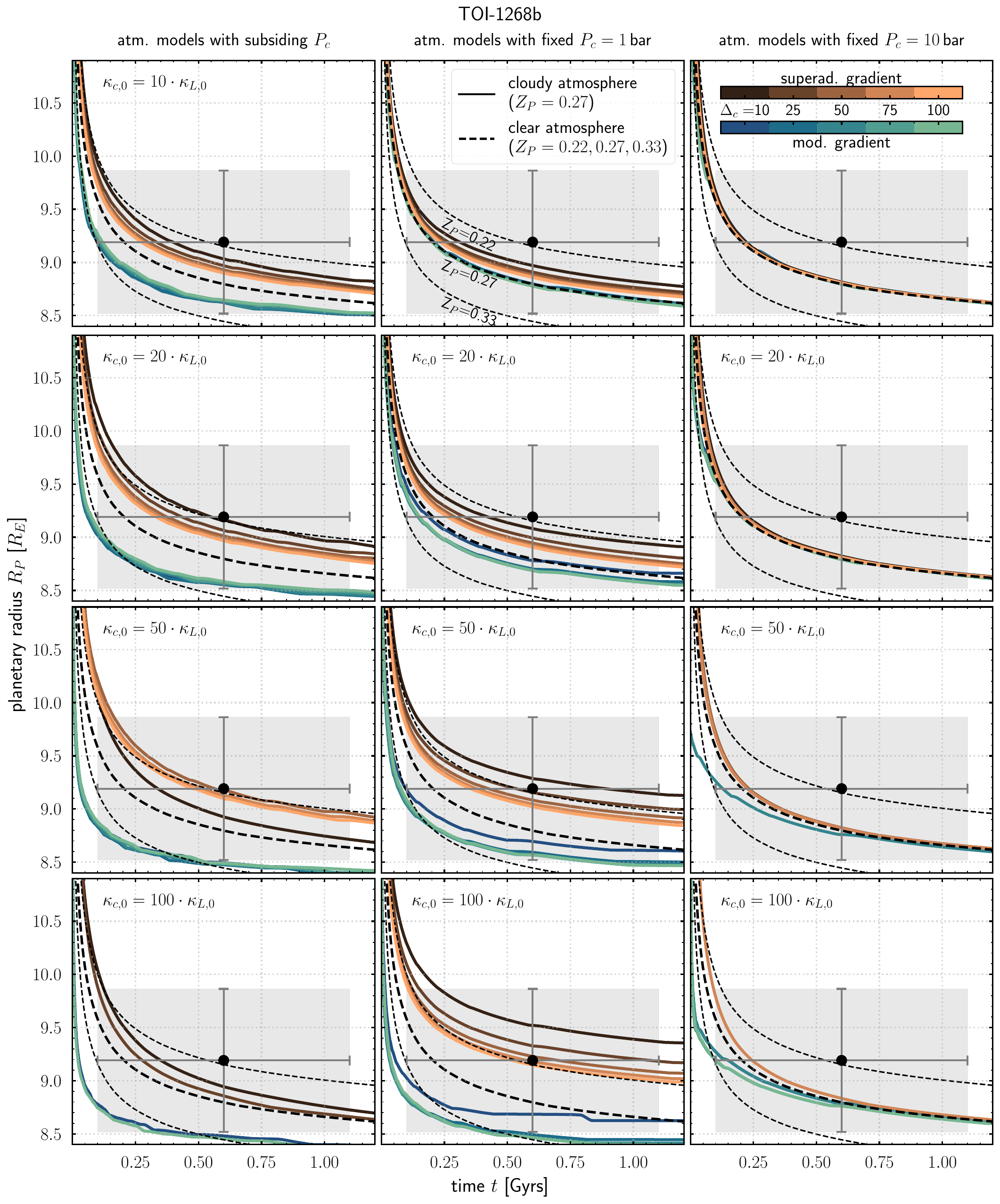}
    \caption{Radius evolution curves for TOI-1268b with subsiding $P_\mathrm{c}$ (left column) using \ch{MgSiO3} as possible condensate in the deeper atmosphere and fixed $P_\mathrm{c}=1,10\,$bar (middle and right column). The evolution model with subsiding  $P_\mathrm{c}$ has a greater impact on the radius evolution than models with a fixed cloud deck. The grey box in each panel represents the uncertainty in measured radius and age. Each panel compares the radius evolution without clouds (dashed curves) and with clouds (solid coloured) for a set of $\kappa_\mathrm{c,0}$ and $\Delta_\mathrm{c}$. The cloudy models are calculated for $Z_\mathrm{P}=0.27$, the corresponding clear model is shown in thick dashed black. Each panel shows the results of the modified gradient (blue tones) and the superadiabatic gradient (orange tones). Additionally, we show in thin dashed black the results with a clear atmosphere for $Z_\mathrm{P}=0.22, 0.33$.}
    \label{fig:TOI1268b_evol_overview}
\end{figure*}

\begin{figure*}
	 \includegraphics[width=0.99\textwidth]{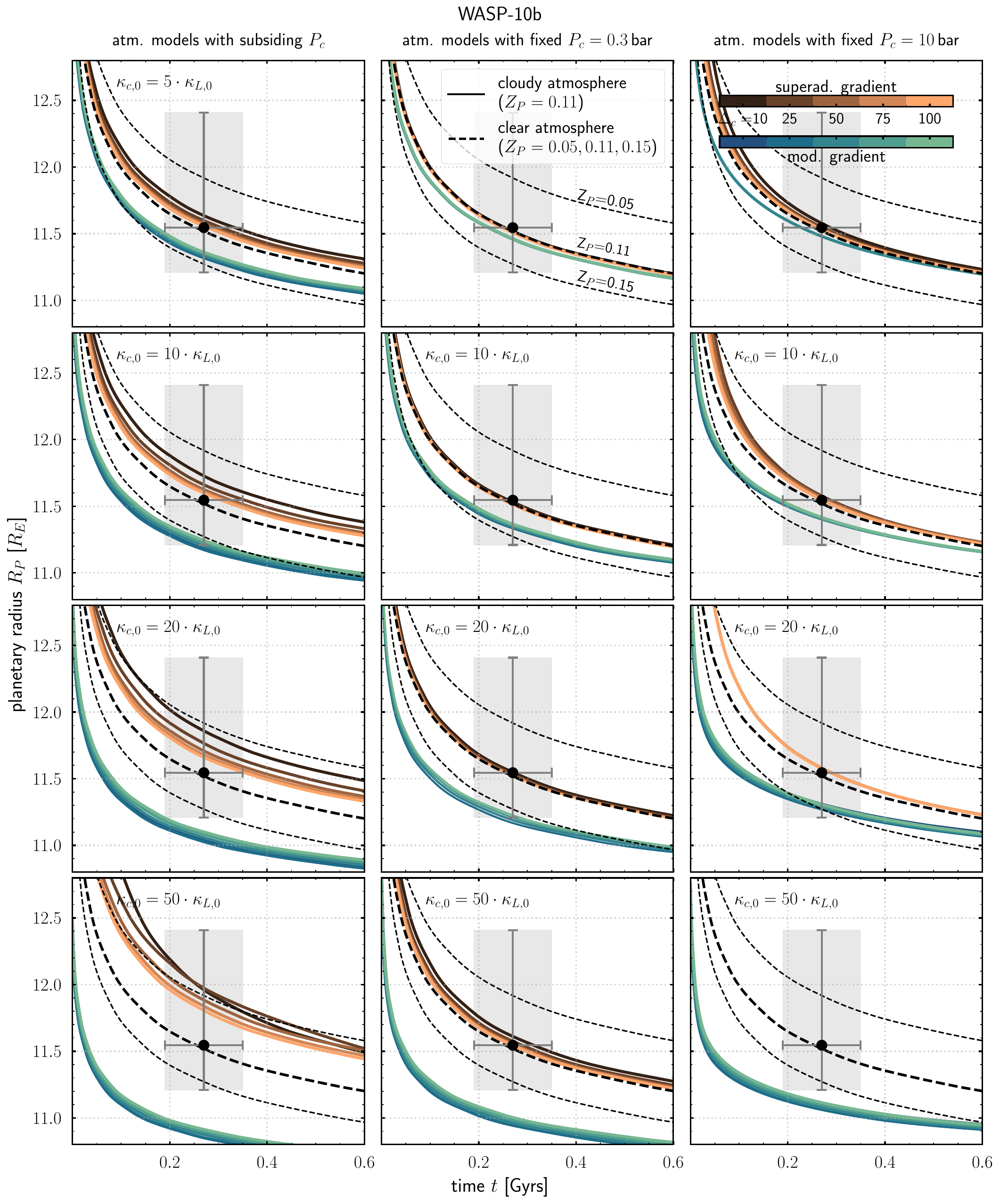}
    \caption{Same as Fig.~\ref{fig:TOI1268b_evol_overview}, but for WASP-10b. The first column shows the results that include the dynamic model to calculate $P_\mathrm{c}$ using \ch{MgSiO3} as the possible condensate, while the second uses the fixed location at $P_\mathrm{c}=0.3\,$bar. The cloudy evolution curves have been calculated for $Z_\mathrm{P}=0.11$, the corresponding clear model is shown in thick dashed black. Additionally, we show in thin dashed black the results with a clear atmosphere for $Z_\mathrm{P}=0.05, 0.15$. Each panel shows the results of the modified gradient (blue tones) and the superadiabatic gradient (orange tones). The atmospheric profiles of Fig.~\ref{fig:WASP10b_PT} are input for the evolution curves of the second row. In cases where the evolution curves for the complete set of parameters are not shown, for example, $\kappa_\mathrm{c,0}=50\cdot \kappa_\mathrm{L,0}$ at $P_\mathrm{c}=10\,$bar, it indicates that we did not obtain a numerical result.} 
    \label{fig:WASP10b_evol_overview}
\end{figure*}

Now we present our main results in Fig.~\ref{fig:TOI1268b_evol_overview} and Fig.~\ref{fig:WASP10b_evol_overview}. Both figures display radius evolution curves for TOI-1268b and WASP-10b, respectively. Here, we investigate the impact of different atmospheric pressure-temperature conditions, including clouds as an additional opacity source, on the long-term radius evolution. We perform a parameter study for both planets, varying the main parameter of our atmosphere model in each panel. \\
As visual anchor points, we plot the observational uncertainty in planetary radius and stellar age (grey area in the background).  We calculate the evolution curves of the clear and various cloudy radius evolution curves for a fixed $Z_\mathrm{P}$ (TOI-1268b: $Z_\mathrm{P}=0.27$ (with $Z_\mathrm{env}=0.015$, $M_\mathrm{core}=24\,M_\mathrm{E}$), WASP-10b: $Z_\mathrm{P}=0.11$ (with $Z_\mathrm{env}=0.015$, $M_\mathrm{core}=100\,M_\mathrm{E}$) to compare within a set of models. Here, $Z_\mathrm{env}=0.015$ is the protosolar value \citep{Lodders2003} which we define as $1\times\,$solar metallicity or [M/H]=0. As we have seen before in Section~\ref{sec:radiusevolution_clear} and Fig.~\ref{fig:TOI1268b_evol_overview}, the thermal evolution is heavily dependent on the bulk metallicity of the planet $Z_\mathrm{P}$. Additionally, we show reference models with a clear, non-cloudy atmosphere for different $Z_\mathrm{P}$ (thin black dashed lines).
As we are interested in the influence of the different parameters of the atmosphere model on the radius evolution, we vary the grey cloud opacity normalisation $\kappa_\mathrm{c,0}$ in each row of the matrix, and within each panel we vary the cloud deck thickness $\Delta_\mathrm{c}=10-100$ where a small number describes a (geometrically) thicker cloud deck. Furthermore, for each set of $\left(\kappa_\mathrm{c,0} /\Delta_\mathrm{c} \right)$ we vary the gradient, see Section~\ref{sec:atm}. The atmospheric profiles from the Heng model may result in a superadiabatic $P$-$T$ gradient in the atmosphere (orange tones) whereas we modify the $P$-$T$ profile so that we inhibit the superadiabaticity (blue tones). 
We are particularly interested in two different options for the long-term evolution of the atmospheric model. The first column shows the results for a \textit{dynamic} cloud deck position (subsiding $P_\mathrm{c}$) during evolution, which changes the location in the atmosphere of the added longwave opacity, see Section~\ref{sec:atm_variability}. The second and third columns show the results for the evolution curves for a fixed cloud base pressure $P_\mathrm{c}$. \\
We detected the following trends for WASP-10b and TOI-1268b: 
\begin{enumerate}
\item
There is a clear influence of the different cloud gradients on the planets thermal evolution. The cloudy atmospheres with the modified and the superadiabatic cloud gradient separate into two bundles of evolution curves for all parameter sets. The atmosphere models with the superadiabatic gradient (orange) slow down the cooling, keeping the planet hotter than in the clear case. Contrary, the curves with the modified non-superadiabatic gradient (blue) accelerate the cooling. Especially for very early timescales, the cloud decks enhance the cooling and fuel the rapid contraction of the planet. For example, for TOI-1286b (Fig.~\ref{fig:TOI1268b_evol_overview}), this is apparent for $t<0.25\,$Gyrs. In further evolution, this rapid cooling stops and the planet shrinks more slowly ($t>0.25\,$Gyrs). The findings of rapid cooling are consistent with the previous work of \citet{Kurosaki2017} for Uranus. However, our results are based on a colder adiabat for a given $T_{\mathrm{int}}$ due to our modification of the cloud model and may not directly represent atmospheric physics as in \citet{Kurosaki2017}.
\item Variation in the thickness of the cloud deck thickness $\Delta_\mathrm{c}$ has different effects:
In the case of the non-modified, superadiabatic cloud gradient (orange), we see that the smaller $\Delta_\mathrm{c}$ is, the larger is the effect of keeping the planet hot, as the greenhouse effect comes more into play (smaller $\Delta_\mathrm{c}$ equals a larger cloud deck thickness). The behaviour is systematic. 
In the case of the modified gradient (blue), for the modified, non-superadiabatic evolution curves and the subsiding $P_\mathrm{c}$, the (geometrical) thickness has a minor influence. 
\item With larger cloud opacity normalisation $\kappa_\mathrm{c,0}$, the effects of the above points are even more pronounced. For models with superadiabatic cloud gradient, the heat is trapped more efficiently, and the cooling slows down more enhanced. In addition, the influence of the thickness of the cloud deck on the evolution of the radius increases as the set of respective evolution curves spreads more. For the modified cloud gradient, the larger $\kappa_\mathrm{c,0}$, the more pronounced is the cooling of the very young planet, and the difference from the clear case becomes larger. 
For WASP-10b (Fig.~\ref{fig:WASP10b_evol_overview}), the effect of the cloud deck thickness behaves systematically for both gradients, leading to a light spread of evolution curves for the modified cloud gradient and a larger spread for the superadiabatic gradient compared to the clear case. 
For TOI-1268b (Fig.~\ref{fig:TOI1268b_evol_overview}), the effect of an enhanced systematic influence of the cloud deck thickness with larger cloud opacity is given for the superadiabatic gradient, but not for all parameter combinations of the models with the modified gradients.
\item 
Lastly, we want to focus on the effects of a variable cloud deck location versus a fixed cloud deck location during the long-term evolution, see Subsection~\ref{sec:atm_variability}. 
For TOI-1268b, the first column takes into account the variability of the cloud deck location. The second and third columns keep the cloud deck fixed at $P_\mathrm{c}=10\,$bar and $P_\mathrm{c}=10\,$bar, respectively. The fixed cloud deck at $P_\mathrm{c}=10\,$bar has a minor influence compared to the other two columns and the clear case. Only for $t<0.25\,$ Gyrs is the effect of faster cooling apparent for models with a modified cloud gradient.
Interestingly, the influence of the variable cloud deck location compared to the fixed cloud deck at $P_\mathrm{c}=1\,$bar depends on the cloud deck opacity. For $\kappa_\mathrm{c,0}=(10,20)\,\kappa_\mathrm{L,0}$, the effect of the variable cloud deck is greater. The effect reverses for $\kappa_\mathrm{c,0}=(50,100)\,\kappa_\mathrm{L,0}$, where the effect is larger for the fixed cloud deck location.
For WASP-10b, the effect of a variable cloud deck location compared to a fixed cloud deck location is greater for all $\kappa_\mathrm{c,0}$.
\end{enumerate}
Our aim is to investigate the effects and impact of atmosphere models with and without clouds on the long-term evolution, which is ultimately important to take into account to determine the planets' bulk metallicity. Therefore, we estimate the effect on $Z_\mathrm{P}$ as follows:\\
We find for TOI-1268b a total heavy element mass of approximately $10-18\,M_\mathrm{E}<M_\mathrm{Z}<27-35\,M_\mathrm{E}$ with a clear atmosphere using CD21 ($Z_\mathrm{P}\approx0.1-0.3$ with CD21, $Z_\mathrm{P}\approx0.2-0.38$ with SCvH95) within the observational error bars. We find that including a cloudy atmosphere model for a specific $Z_\mathrm{P}$ can result in similar clear evolution curves for $\approx Z_\mathrm{P}\pm 0.05$ when taking into account a variable cloud deck position during the planet's long-term evolution. For the fixed cloud deck position at $P_\mathrm{c}=1\,$bar and high $\kappa_\mathrm{c,0}=(50,100)\,\kappa_\mathrm{L,0}$ we find that the curves equal $\approx {Z_\mathrm{P}}_{-0.05}^{+0.10}$.\\
For WASP-10b, we find a total heavy element mass of approximately $M_\mathrm{Z}<140-160\,M_\mathrm{E}$ with a clear atmosphere using CD21 ($Z_\mathrm{P}\approx0-0.15$ with CD21, $Z_\mathrm{P}\approx0.01-0.17$ with SCvH95). We find that including a cloudy atmosphere model for a specific $Z_\mathrm{P}$ can result in similar clear evolution curves for $\approx {Z_\mathrm{P}}_{-0.03}^{+0.06} $ (for smaller $\kappa_\mathrm{c,0}=(5,10)\,\kappa_\mathrm{L,0}$) when taking into account a variable cloud deck position during the planet's long-term evolution. For the variable cloud deck position and high $\kappa_\mathrm{c,0}=(20,50)\,\kappa_\mathrm{L,0}$ we find that the curves equal $\approx {Z_\mathrm{P}}_{-0.06}^{+0.10}$.\\
We argue that the atmosphere model is therefore a source of degeneracy while determining the planets' metallicity and suggest taking into account the atmospheric $P$-$T$ structure during the planet's evolution.
This is especially important when the observational uncertainties become smaller with upcoming missions.\\
Note that we have not applied a statistical approach. We computed the thermal evolution of individual models ($\approx$ 200), adjusting the parameter space of the atmosphere model rather than accounting for observational uncertainties such as the mass of the planet. Regarding the interplay between observational and theoretical uncertainties, \citet{Muller2020} discovered that theoretical uncertainties can be comparable to or even exceed observed uncertainties. However, we have not specifically explored this dynamic within our set of planets in this particular study.

\section{Discussion}
\label{sec:discussion}

\begin{figure}
	\includegraphics[width=0.99\columnwidth]{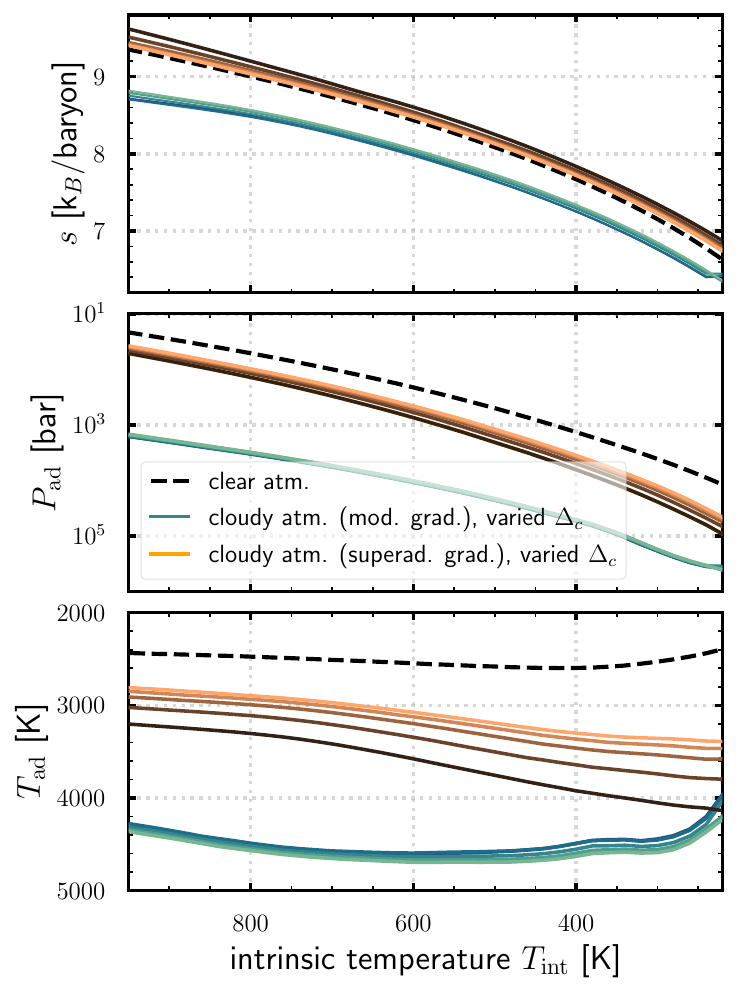}
    \caption{Exemplary, we plot values of the adiabatic temperature $T_{\mathrm{ad}}$, pressure $P_{\mathrm{ad}}$ as the radiative-convective boundary (RCB) values (lower panels) and resulting entropy $s$ (upper panel) for the evolution of $T_{\mathrm{int}}$ during the thermal evolution of WASP-10b for different model atmospheres. The parameters of the atmosphere models correspond to $\kappa_\mathrm{c,0}=10\,\kappa_\mathrm{L,0}$ and subsiding $P_\mathrm{c}$(\ch{MgSiO3}), see Fig.~\ref{fig:WASP10b_evol_overview}. The dashed line indicates the reference clear atmospheric case. We calculate the RCB by comparing the local and adiabatic gradients, see Eq.~\ref{eq:cloudopacity}.}
    \label{fig:RCB}
\end{figure}

Matching the observed radius, mass, and age of the planet with numerical models naturally leads to a number of degeneracies of the resulting interior structure. In this work, we show the impact of H/He EoS, the distribution of metals, and the uncertainty of the observable parameters (planetary mass and radius) on the $T_{\mathrm{int}}$-$Z_{\mathrm{env}}$ phase space of the two warm giant planets. We highlight the additional uncertainty that comes into play with a cloudy atmosphere model which we focus on in this paper. 
Second, by comparing the radius evolution curves using different atmospheric models, we confirm the results of previous publications that atmospheric conditions have an impact on the planets' thermal evolution \citep[e.g.][]{Vazan2013, Kurosaki2017}. We refine this result by inserting cloud decks into the atmosphere, looking at the effects of the $P$-$T$ structure on the planet's evolution. 
We want to point out and discuss the obtained results in the following paragraphs.\\
First, there are several caveats regarding the atmosphere model used in our study:
The real atmosphere is much more diverse than represented by our 1D averaged atmospheric model. Pressure and temperature vary in 3D and in time.  Most of the hot Jupiters are tidally locked, with the effect of day/night side temperature gradients with a varied heat distribution depending on $T_{\mathrm{eq}}$. Furthermore, an enhanced atmospheric metallicity pushes the $P$-$T$ profile to higher temperatures \citep[e.g.][]{Drummond2018, Fortney2006}, which may underestimate the cloud base. Observational measurements, such as phase curve observations and transmission spectra, can only give estimates of the physical conditions, e.g., resulting in broad assumptions on temperature and pressure. 
The occurrence of clouds may depend on stellar irradiation: By analysing the spectra of a sample of irradiated planets, \citet{Estrela2022} find a group of atmospheres with a trend from cloudy/hazy to clear, in the range of $T_{\mathrm{eq}}=500-1500\,$K. We infer that both planets of this study with $T_{\mathrm{eq}}\approx900\,$K could possibly accommodate clouds. We note that clouds may additionally not occur as one permanent cloud deck, but patchy. 
Furthermore, within the atmosphere model used, we consider the effect of clouds as purely absorbing without scattering effect. Adding an extra cloud opacity in the longwave then leads to a warming Greenhouse effect, which places an upper limit on our considerations. In addition, we do not consider the possible interaction of cloud decks. We note that there must be enough material to condense out in the deep atmosphere regions to build up a cloud deck, and that rainout may play a role~\citep[e.g.][]{Mbarek2016}. \\
Second, an important aspect of our coupled atmosphere-interior-evolution model is the connection to the deep interior at the radiative-convective boundary. The location of the RCB determines the interior adiabat and hence influences the planets' thermal evolution. At the RCB, temperature fluctuations may not be as strong as in the upper atmosphere, influencing the $P$-$T$ profile in a minor way. On the other hand, the chosen atmosphere model does greatly impact the RCB and hence the cooling behaviour. In Fig.~\ref{fig:RCB}, we show the development of the entropy of the interior and the corresponding $T_{\mathrm{ad}}$, $P_{\mathrm{ad}}$ during the planet's evolution, plotted over the intrinsic temperature. For this example, we use the results of Fig.~\ref{fig:WASP10b_evol_overview} (model parameter: $\kappa_\mathrm{c,0}=10\,\kappa_\mathrm{L,0}$ and subsiding $P_\mathrm{c}$(\ch{MgSiO3}). The reference case with the clear atmosphere is shown in black dashed. In line with the results of the radius evolution, the faster cooling of the atmospheric models with the modified gradient (blue curves) is due to the lower entropy of the adiabat. On the contrary, the slower cooling is due to a higher entropy. The adiabatic pressure $P_{\mathrm{ad}}$ is higher than in the clear case for a specific $T_{\mathrm{int}}$ value. The adiabatic temperature $T_{\mathrm{ad}}$ is  $\approx 1000\,$K higher for the atmospheric model with the modified gradient. In Fig.~\ref{fig:WASP10b_PT} (a), (c) and (d), the position of the RCB is shown as a circle.  The wobble in $T_{\mathrm{ad}}$ for $T_{\mathrm{int}}<500\,$K stems from the calculation of the RCB where we compare the gradients, Eq.~\ref{eq:RCB}, using the adiabatic gradient from the EoS tables. We want to point out that, regardless of the physical phenomena, the RCB is impacting the thermal long-term evolution of the planet. \\ 
Third, looking at the impact of observational uncertainties versus the uncertainty given due to the atmosphere model with and without clouds, we note that the results due to different atmospheric models for nearly all parameters lie in the uncertainty range of radius and age. Better constraints on planetary radius, mass and stellar age are needed to characterise the planets and to narrow down the parameter space, such as aimed at by space missions, e.g. PLATO (\citet{Rauer2014}).\\
In general, the findings of this study confirm that the atmosphere plays a crucial role for the radius evolution of a planet. Further, time-variable cloud decks may have a significant impact on the contraction of the planet, adding substantially to the model degeneracy when coupling atmosphere-interior-thermal evolution models for warm giant planets. We suggest taking into account the time-variability of the deep atmosphere during the long-term evolution of gas giants.  \\

\section{Summary and Conclusions}
In this paper, we explore the impact of cloud decks on the $T_{\mathrm{int}}$-$Z_{\mathrm{env}}$ phase space and radius evolution for two young warm gas giants, WASP-10b and TOI-1268b. The main focus of this paper was to extend the previous work on the effect of clouds on the thermal evolution of irradiated gas planets. We focus on cloud decks in the deep atmosphere. Ultimately, this may help to constrain the metal content of the planet.\\ 
This work is based on the previous work by~\citet{Poser2019}. We used a conventional three-layer model consisting of a core, an adiabatic envelope, and a radiative atmosphere to model the thermal evolution of the planets. For the pressure and temperature of the atmosphere, we used a semi-analytical model with grey opacities. To account for cloud decks in the atmosphere, we added a purely absorbing cloud deck resulting in a warming effect for the deep atmosphere.  The cloud deck is described as an additional grey opacity, added to the long-wave opacity. We assume the cloud deck to be formed where there is an intersection with a condensation curve of a cloud forming species. Within this model, it is possible to investigate general trends in the atmospheric temperature structure for the (thermal) radius evolution. To illustrate the impact of cloud decks, we compare several atmospheric model setups during the planets thermal evolution. We summarise our main findings as follows: 
\begin{enumerate}
    \item The additional infrared opacity warms the atmosphere beneath. For the thermal radius evolution of the planet, that leads to either a slower/faster cooling than in the clear case. The specific outcome hinges on the choice of cloud gradient, whether superadiabatic or not, see Figs.~\ref{fig:TOI1268b_evol_overview} and~\ref{fig:WASP10b_evol_overview}.
    \item When comparing the effect of a fixed cloud base level versus a dynamic cloud base level during the planets' thermal evolution, there is a slight dependence on the cloud opacity for TOI-1268b: For smaller cloud opacities ($\kappa_\mathrm{c,0}<20\,\kappa_\mathrm{L,0}$), we see an enhanced behaviour that results in a faster/slower cooling behaviour (depending on the cloud gradient used) for the dynamic cloud base level. For WASP-10b, the dynamic cloud base shows a stronger effect than the fixed cloud deck case for all cloud opacities.
    \item We demonstrate that atmospheric models including deep clouds can lead to a degeneracy in predicting the planets' bulk metallicity. For the Jupiter-mass WASP-10b, we find a possible span of $\approx {Z_\mathrm{P}}_{-0.06}^{+0.10}$. For the Saturn-mass TOI-1268b, this range extends to $\approx {Z_\mathrm{P}}_{-0.05}^{+0.10}$.
\end{enumerate}    
Additionally, we find that the choice of the EoS (CD21 vs. SCvH95) plays a more significant role in affecting the less dense warm Saturn TOI-1268b compared to the denser warm Jupiter WASP-10b. When comparing the impact of the atmosphere model on the radius evolution of both planets, we find that quantifying the results with respect to the planets' density is not feasible. However, it is likely that such quantification could be achieved with a larger sample size, which we did not undertake. 
Our findings are based on a non-statistical approach, calculating individual models, solely varying the parameter of the atmosphere model. The results can be seen as a first step towards a more sophisticated modelling approach, including the observational uncertainties.\\
However, this study is important in the context of modelling the interior properties of giant planets. It highlights the importance of coupled interior, atmosphere, and thermal evolution models and underlines the role of atmospheric chemistry and cosmochemistry. \\
Overall, we stress the importance of reducing not only the observational uncertainties in planetary radius and mass but also the uncertainty in stellar age as a proxy for the planets' age, supporting the work of, e.g., \citet{Muller2020}, \citet{Muller2023}. Additionally, to further inform planetary formation models, interior models require the planets' atmospheric metallicity as input parameter to point a proper picture. Missions such as \textit{JWST}, \textit{TESS} and the upcoming ESA \textit{ARIEL} mission will address these points, aiming at reducing observational error bars for radius and stellar age as well as providing values of the planets' atmospheric metallicity.

\section*{Acknowledgements}
We thank N. Nettelmann for valuable input to the manuscript, and for data included in the $\gamma(T_\mathrm{eq})$-fit. We thank L. Scheibe, M. Schörner, C. Kellermann, and S. Schumacher for helpful discussions. This work is supported by the DFG project SPP-1992 "Exploring the Diversity of Extrasolar Planets". We thank the anonymous referee for helpful feedback that greatly improved the manuscript.

\section*{Data Availability}
The data that support the findings of this study are available from the corresponding author upon reasonable request.
%The inclusion of a Data Availability Statement is a requirement for articles published in MNRAS. Data Availability Statements provide a standardised format for readers to understand the availability of data underlying the research results described in the article. The statement may refer to original data generated in the course of the study or to third-party data analysed in the article. The statement should describe and provide means of access, where possible, by linking to the data or providing the required accession numbers for the relevant databases or DOIs.


\bibliographystyle{mnras}
\bibliography{library} 

%%%%%%%%%%%%%%%%% APPENDICES %%%%%%%%%%%%%%%%%%%%%
\appendix

\section{Planetary data used for the $\gamma$-fit formula }
\label{sec:appendixgamma}

\begin{landscape}
\begin{table}    
    \centering
    \begin{tabular}{@{}|l|l|l|l|l|l|l|l|l|l|@{}}
    \hline
        Planet & gravity [cm/s$^2$]	& $T_{\mathrm{eq}}$ [K] & $A_B$ & [M/H] & $T_{\mathrm{iso}}$ [K]& $\gamma$ & $\kappa_{\mathrm{L}}$ [m$^2$/kg]& $\kappa_{\mathrm{S}}$ [m$^2$/kg]  & \\   \hline \hline
        GJ1214b & 880.0 & 544.0 & 0.1 & 1 & 1000.0           & 0.050 & 1.0e-03 & 5.00e-05  &  \citet{Miller-Ricci2010}  \\ \hline
        GJ436b & 1270.0 & 655.0 & 0.1 & 1 & 1100.0          & 0.060 & 1.0e-03 & 6.00e-05  & \citet{Morley2017}\\ \hline
        generic & 2570.0 & 857.0 & 0.1 & 1 & 1350.0         & 0.100 & 1.0e-03 & 1.00e-04  & \citet{Fortney2007}\\ \hline
        WASP-39b & 430.4 & 1088.4 & 0.1 & 1 & 1546.0        & 0.154 & 5.0e-04 & 7.70e-05  &  \citet{Wakeford2018}\\ \hline
        HD189733b & 2120.0 & 1169.0 & 0.1 & 1 & 1700.0      & 0.140 & 1.0e-03 & 1.40e-04  & \citet{Madhusudhan2009} \\ \hline
        HD189733b & 2120.0 & 1169.0 & 0.1 & 1 & 1500.0      & 0.260 & 1.0e-03 & 2.60e-04 & \citet{Fortney2010a} \\ \hline
        HD189733b & 2120.0 & 1169.0 & 0.1 & 1 & 1520.0      & 0.240 & 1.0e-03 & 2.40e-04  & \citet{Fortney2010a}\\ \hline
        HD189733b & 2120.0 & 1169.0 & 0.1 & 1 & 1600.0      & 0.190 & 1.0e-03 & 1.90e-04  & \citet{Heng2012}\\ \hline
        fid. planet & 1500.0 & 1267.0 & 0.0 & 1 & 1602.0    & 0.243 & 7.0e-04 & 1.70e-04 & \citet{Jin2014}\\ \hline
        HD209645b & 965.0 & 1402.0 & 0.1 & 1 & 1720.0       & 0.330 & 1.0e-03 & 3.30e-04  & \citet{Fortney+2005}\\ \hline
        HD209458b & 924.7 & 1479.5 & 0.0 & 1 & 1732.0       & 0.357 & 7.0e-04 & 2.50e-04  &\citet{Fortney+2005}\\ \hline
        HD209458b & 924.7 & 1479.5 & 0.0 & 1 & 1796.0       & 0.300 & 2.0e-04 & 6.00e-05 &\citet{Fortney+2005}\\ \hline
        fid. planet  & 1500.0 & 1577.0 & 0.0 & 1 & 1995.0   & 0.250 & 6.0e-04 & 1.50e-04  & \citet{Jin2014}\\ \hline
        HAT-P-13b & 1286.0 & 1605.0 & 0.1 & 1 & 2000.0      & 0.300 & 1.0e-03 & 3.00e-04 & \citet{Kramm2012}\\ \hline
        HD149026b & 1697.7 & 1629.0 & 0.1 & 1 & 1900.0      & 0.424 & 1.0e-03 & 4.24e-04 & \citet{Fortney2006}\\ \hline
        WASP-103b & 1574.0 & 2444.0 & 0.1 & 1 & 2700.0      & 0.600 & 1.0e-03 & 6.00e-04 &  \citet{Kreidberg2018a} \\ \hline
        fid. planet & 1500.0 & 2777.0 & 0.0 & 1 & 2905.0    & 0.717 & 6.0e-03 & 4.30e-03  & \citet{Jin2014}\\ \hline
        KELT-9b & 1902.5 & 4050.0 & 0.0 & 1 & 4100.0 & 0.900 & 1.0e-03 & 9.00e-04   & \citet{Mansfield2020, Fossati2020}\\ \hline\hline
    \end{tabular}
    \caption{Planetary parameter used for the $\gamma$-fit shown in Fig.~\ref{fig:fitformula_na}. Note that $\kappa_\mathrm{L}$ here refers to the notation of the clear atmosphere model by \citet{Guillot2010}. It corresponds to $\kappa_{\mathrm{L,0}}$ in the notation used by \citet{Heng2012}.  }
    \label{tab:fitformula_gamma}
\end{table}

\begin{table}
    \centering
\begin{tabular}{@{}|l|l|l|l|l|l|l|l|l|l|@{}}
\hline
        Planet & gravity [cm/s$^2$]	 & $T_\mathrm{eq}$ [K] & $A_B$ & [M/H] & $T_{\mathrm{iso}}$ [K] & $\gamma$ & $\kappa_{\mathrm{L}}$ [m$^2$/kg]& $\kappa_{\mathrm{S}}$ [m$^2$/kg]&\\ \hline\hline
    WASP-10b & 6915.9   & 869.0 & 0.3 & 1       & 1360.0 & 0.147 & 1.4e-03 & 2.00e-04    &  \citet{Fortney2007}\\ \hline
    WASP-39b & 430.4    & 1088.4 & 0.1 & 10     & 1835.0 & 0.065 & 1.3e-03 & 8.27e-05   & \citet{Molliere2015}\\ \hline
    HD149026b & 1697.7  & 1629.0 & 0.1 & 3      & 2000.0 & 0.320 & 1.0e-03 & 3.20e-04    & \citet{Fortney2006}\\ \hline
    HD149026b & 1697.7  & 1629.0 & 0.1 & 10     & 2200.0 & 0.202 & 1.0e-03 & 2.02e-04    & \citet{Fortney2006}\\ \hline
\hline
\end{tabular}
\caption{Planets \textit{not used} for the $\gamma$-fit, but displayed in Fig.~\ref{fig:fitformula_na}. Here, the metallicities are larger than 1x solar abundance, or it is $A_B>0.1$. Note that $\kappa_\mathrm{L}$ here refers to the notation of the clear atmosphere model by \citet{Guillot2010}. It corresponds to $\kappa_{\mathrm{L,0}}$ in the notation used by \citet{Heng2012}.}
\label{tab:fitformula_gamma_unused}
\end{table}
\end{landscape}

% Don't change these lines
\bsp	% typesetting comment
\label{lastpage}
\end{document}